\documentclass[onecolumn,draftclsnofoot,12pt]{IEEEtran}

\usepackage{booktabs}
\usepackage{amsmath}
\usepackage{amsthm}
\usepackage{multirow}
\usepackage{color,cite}
\usepackage{subfigure}
\usepackage{mathrsfs}
\usepackage{amssymb}
\usepackage{bm}
\usepackage{graphicx}
\usepackage{amsfonts}
\usepackage{algorithm}
\usepackage{algorithmic}
\usepackage{epstopdf}

\newcommand{\tabincell}[2]{\begin{tabular}{@{}#1@{}}#2\end{tabular}}

\begin{document}

\bibliographystyle{IEEEtran}

\title{\huge{Deep Learning Empowered Type-II Codebook:\\New Paradigm for Enhancing CSI Feedback}}

\author{
\vspace{-1mm}
Ke Ma, \emph{Graduate Student Member, IEEE}, Yiliang Sang, Yang Ming,\\ Jin Lian, Chang Tian, Zhaocheng Wang, \emph{Fellow, IEEE}
\thanks{This work was supported in part by the National Key R\&D Program of China under Grant 2018YFB1801102, and in part by Huawei Research Fund. Part of this work has been submitted to IEEE GLOBECOM 2023. \emph{(Corresponding author: Zhaocheng Wang)}} %
\thanks{K.~Ma, Y. Sang, Y. Ming and Z. Wang are with Beijing National Research Center for Information Science and Technology, Department of Electronic Engineering, Tsinghua University, Beijing 100084, China. Y. Ming and Z. Wang are also with Shenzhen International Graduate School, Tsinghua University, Shenzhen 518055, China (E-mails: ma-k19@mails.tsinghua.edu.cn, mingy20@mails.tsinghua.edu.cn, sangyl19@mails.tsinghua.edu.cn, zcwang@tsinghua.edu.cn).} %
\thanks{J.~Lian and C.~Tian are with Huawei Technologies Company, Ltd., Beijing 100095, China (E-mails: lianjin7@huawei.com, tianchang7@huawei.com).}
% \thanks{J.~Cheng is with the School of Engineering, The University of British Columbia, Kelowna, BC V1V 1V7, Canada (E-mail: julian.cheng@ubc.ca).} %
\vspace{-18mm}
}

\maketitle

\begin{abstract}
Deep learning based channel state information (CSI) feedback in frequency division duplex systems has drawn much attention in both academia and industry.
In this paper, we focus on integrating the Type-II codebook in the beyond fifth-generation (B5G) wireless systems with deep learning to enhance the performance of CSI feedback.
In contrast to its counterpart in Release~16, the Type-II codebook in Release~17 (R17) exploits the angular-delay-domain partial reciprocity between uplink and downlink channels and selects part of angular-delay-domain ports for measuring and feeding back the downlink CSI, where the performance of the conventional deep learning methods is limited due to the deficiency of sparse structures.
To address this issue, we propose the new paradigm of adopting deep learning to improve the performance of R17 Type-II codebook.
Firstly, considering the relatively low signal-to-noise ratio of uplink channels, deep learning is utilized to refine the selection of the dominant angular-delay-domain ports, where the focal loss is harnessed to solve the class imbalance problem.
Secondly, we propose to reconstruct the downlink CSI by way of deep learning based on the feedback of R17 Type-II codebook at the base station, where the information of sparse structures can be effectively leveraged.
Finally, a weighted shortcut module is designed to facilitate the accurate reconstruction, and a two-stage loss function with the combination of the mean squared error and sum rate is proposed for adapting to actual multi-user scenarios.
Simulation results demonstrate that our proposed angular-delay-domain port selection and CSI reconstruction paradigm can improve the sum rate performance by more than $10\%$ compared with the traditional R17 Type-II codebook and deep learning benchmarks.
\end{abstract}


\IEEEpeerreviewmaketitle

\section{Introduction}\label{S1}

Benefitting from the high spectral efficiency, massive multiple-input multiple-output (MIMO) has been regarded as one of the fundamental technologies in the fifth-generation (5G) wireless communications and beyond \cite{ref1, ref2}.
In order to provide efficient precoding for supporting reliable transmissions in multi-user (MU) scenarios, the base station (BS) needs to acquire accurate downlink channel state information (CSI) of each user (UE) \cite{ref3}.
For the time division duplex (TDD) systems, the channel reciprocity can be utilized to obtain the downlink CSI based on uplink channel measurements \cite{ref4}, which is unfortunately not applicable in the frequency division duplex (FDD) systems.
Therefore, each UE is required to feed back its downlink CSI to BS in FDD systems, which could impose overwhelmingly large overhead \cite{ref5}.
% How to reduce the overhead of CSI feedback without severe accuracy loss has become an imperative issue in FDD systems.
How to reduce the overhead of CSI feedback with ensuring high CSI accuracy has become an imperative issue for FDD systems.

Fortunately, CSI in massive MIMO systems usually enjoys the sparse property in both angular and delay domains owing to the limited number of scatterers around the surrounding environment \cite{ref6, ref7}, which provides the feasibility to transform the original CSI into the corresponding sparse space for efficient compression.
Consequently, diverse compressed sensing (CS) based methods have been proposed to obtain the sufficiently accurate CSI estimates based on a small number of sparse bases \cite{ref8, ref9, ref10, ref11}.
%, including orthogonal catching pursuit (OMP), approximate message passing (AMP).
Moreover, numerous studies and field experiments have validated the angular-delay-domain partial reciprocity in FDD systems, i.e., both the uplink and downlink CSI have similar power properties in both angular and delay domains \cite{ref7, ref11, ref12, ref13}, which motivated many technical works to select the strongest angular-delay-domain ports according to the uplink CSI for downlink channel measurements \cite{ref14, ref15}, so that the downlink pilot overhead could be significantly reduced.
However, the limited transmit power of UE usually leads to relatively low signal-to-noise ratios (SNRs) of uplink CSI, making it difficult to select the dominant angular-delay-domain ports accurately.
% Convinced by its advantages, this method has been supported in the Type-II codebook in the latest Release 17 (Rel.~17) communication standards.

% In spite of widespread applications, the conventional CS based methods suffer from a crucial drawback that they heavily rely on pre-determined sparse assumptions and measurement matrices, whose performance may degrade severely if the practical channel is not consistent with these assumptions.
% In spite of widespread applications, the conventional CS based methods suffer from a crucial drawback that they heavily rely on the pre-determined sparse measurement matrix, which is usually non-optimal for practical channels.
In spite of widespread applications, the conventional CS based methods suffer from a crucial drawback that they heavily rely on pre-determined sparse assumptions, whose performance may degrade severely if the practical channel is not consistent with these model assumptions.
% For example, the DFT bases are widely adopted to implement the angular-delay-domain transform, whereas its power leakage may seriously ruin the sparsity and lead to the accuracy deterioration of CSI compression [16], [17].
To address this issue, deep learning has been introduced into CSI feedback, whose data-driven optimization manner could adaptively capture the channel sparsity in practical scenarios for enhancing the feedback performance.
As a pioneering study, CsiNet \cite{ref16} firstly utilized the auto-encoder (AE) to implement the CSI feedback task, which demonstrated its amazing superiority over the conventional CS based methods, especially in the scenarios with high compression ratios.
After that, many studies focused on designing more powerful network architectures to further enhance the feedback performance \cite{ref17, ref18, ref19, ref20, ref21}. Taking practical deployments into consideration, the dedicated quantization methods were integrated with deep learning to reduce the feedback overhead \cite{ref18, ref22, ref23}, and the light-weight model designs have attracted much attention \cite{ref24, ref25, ref26, ref27, ref28}.
By utilizing the angular-delay-domain partial reciprocity, the uplink CSI information was fused in the AE to relieve the burden of UE feeback \cite{ref29, ref30, ref31}.
Furthermore, the precoding design has been combined with deep learning based CSI feedback to improve the sum rate performance in MU scenarios \cite{ref32, ref33, ref34}.

% Although the existing studies originated from CsiNet mainly concentrate on the compression and feedback of the full CSI martix, third Generation Partnership Project (3GPP) agreed the CSI eigenvector feedback as from multiple OFDM subbands as the specification [32].
Although the existing studies originated from CsiNet mainly concentrate on the compression and feedback of the full CSI matrix, the Third Generation Partnership Project (3GPP) adopts the CSI eigenvector feedback in 5G wireless networks \cite{ref35}, where the CSI eigenvectors of multiple subbands in orthogonal frequency division multiplexing (OFDM) systems are jointly compressed and fed back.
% Consequently, exploring efficient CSI eigenvector feedback with deep learning has drawn broad attention.
In \cite{ref36}, EVCsiNet was constructed to exploit the AE architecture for feeding back the CSI eigenvectors.
Besides, the work \cite{ref37} proposed MixerNet to fuse spatial and frequency features for improving the eigenvector feedback accuracy, while the dual-polarized antenna structure was exploited in \cite{ref_add1}.
Moreover, the bi-directional long short-term memory (bi-LSTM) network was adopted in \cite{ref38} to model the eigenvector variations among different subbands.
% PolarDenseNet?

% However, the above methods for CSI eigenvector feedback focus on compressing the whole eigenvectors as in the Release 16 (R16) Type-II codebook, which still imposes considerable overhead under a large number of BS antennas.
However, the state-of-the-art methods for CSI eigenvector feedback only focus on improving the performance of the Release 16 (R16) Type-II codebook, which needs to measure the whole downlink CSI eigenvectors for choosing the sparse bases and imposes considerable downlink pilot overhead.
To tackle this problem, the Release 17 (R17) Type-II codebook further leverages the angular-delay-domain partial reciprocity to select the dominant angular-delay-domain ports based on the uplink CSI information and only measures the corresponding channel coefficients, so that the downlink pilot overhead can be significantly reduced \cite{ref35}.
For the R17 Type-II codebook, different UEs usually select various angular-delay-domain ports for downlink channel measurement, making the sparse structure of the feedback coefficient vector, which is reflected by the correlations among different channel coefficients, vary for diverse UEs.
Since UE is not aware of its corresponding angular-delay-domain ports, the stable sparse structure of the feedback vector might be destroyed, which restricts the performance of deep learning based CSI compression at UE side, since deep learning is not good at extracting the features from non-sparse data \cite{ref40, ref41}.
% Moreover, the simulation analysis based on the 3GPP TR38.901 channel model [38] shows that the standard R17 Type-II codebook could achieve the channel cosine similarity more than $99.2\%$, which again demonstrates that deep learning may be very difficult to achieve a better compression performance.

In contrast to directly adopting deep learning to feed back the CSI information, a new paradigm is proposed to improve the feedback performance of R17 Type-II codebook.
Firstly, considering the relatively low SNR of uplink channel measurement, deep learning is used to capture the sparse structure in the uplink CSI for refining the selection of the dominant angular-delay-domain ports. Nevertheless, the selected angular-delay-domain ports only occupy a small fraction of all the available ports, resulting in the unbalanced classes, and the ports with lower power are difficult to be accurately chosen. To solve these problems, the focal loss is harnessed to balance both the positive/negative classes and the simple/hard samples for simultaneously accelerating model convergence and improving the selection accuracy.
Secondly, we propose to use deep learning to reconstruct the downlink CSI eigenvectors on all angular-delay-domain ports based on the feedback of R17 Type-II codebook at BS side.
Different from the traditional CSI compression at UE side, the BS can place each channel coefficient on its corresponding port as the reconstruction input. As a result, the sparse structure in the angular-delay domain can be sufficiently leveraged.
Finally, since the CSI reconstructed by the feedback information form the Type-II codebook is similar to the perfect CSI, we adopt a weighted shortcut module to implement the fine-tuning on the reconstruction input.
In order to maximize the UE received power and decrease the inter-UE interference simultaneously, a two-stage loss function that integrates the mean squared error (MSE) and sum rate is proposed for adapting to MU scenarios.
Specifically, at the first stage, the MSE loss is applied to guarantee high accuracy of CSI reconstruction.
At the second stage, the deep learning model is optimized by both the MSE loss and the negative sum rate loss for mitigating the inter-UE interference effectively.
Simulation results demonstrate that our proposed deep learning based port selection and CSI reconstruction methods can improve the sum rate performance by more than $10\%$ over its traditional R17 Type-II codebook and deep learning benchmarks.

The main contributions of this paper are recapped as follows.
\begin{itemize}
\item{Deep learning is firstly proposed to enhance the accuracy of the angular-delay-domain port selection in the R17 Type-II codebook, where the focal loss is harnessed to adaptively weight the loss function for accelerating model convergence and improving the selection accuracy.}
    % a dual-band fusion approach to integrate low-frequency CSI with mmWave low-overhead measurement to predict the optimal mmWave beam in the HetNet and reduce the huge overhead of both mmWave BS selection and beam training. Furthermore, we propose to adopt deep learning to extract the complex relationships between low-frequency CSI and mmWave low-overhead measurement, where the attention mechanism is leveraged to adaptively weight the features for enhancing the prediction accuracy.}
\item{The downlink CSI eigenvectors on all angular-delay-domain ports are proposed to be reconstructed based on the feedback information from the R17 Type-II codebook at BS side, so that the sparse structure can be fully leveraged. Besides, we propose a weighted shortcut module to enhance the reconstruction accuracy, and a two-stage loss function including the criteria of both MSE and sum rate is proposed to improve the reconstruction performance in actual MU scenarios.}
    %adopt the UE-specific high-quality mmWave wide beams predicted by the low-frequency CSI as the mmWave low-overhead measurement. We also design two criteria for the wide beam selection. The fixed-number criterion selects the fixed number of the max-probability wide beams, while the sum-probability criterion selects the max-probability wide beam set whose probability sum is larger than a predefined threshold, which can adapt to different confidences of the mmWave wide beam prediction for diverse UE.}
\end{itemize}

The paper is organized as follows. Section~\ref{S2} presents the system model and the R17 Type-II codebook. Our proposed deep learning based port selection and CSI reconstruction methods are detailed in Sections~\ref{S3} and \ref{S4}, respectively. Section~\ref{S5} provides the simulation results and performance analysis. Our conclusions are drawn in Section~\ref{S6}.

The following notational conventions are applied in this paper.
% $\mathbb{N}^+$ is the set of positive integers.
$\mathbb{R}^{m\times n}$ and $\mathbb{C}^{m\times n}$ are the $m\times n$ real and complex spaces, respectively.
% $\mathbb{C}^{m\times n}$ represents the $m\times n$ complex space.
Boldface capital and lower-case letters represent matrices and vectors, e.g., $\bm{A}$ and $\bm{a}$, while calligraphic capital letters stand for sets, e.g., $\mathcal{A}$.
The $i$-th row and $j$-th column in matrix $\bm{A}$ are represented as $\bm{A}{[i, :]}$ and $\bm{A}{[:, j]}$, respectively, and the corresponding element in the $i$-th row and $j$-th column is $\bm{A}{[i, j]}$.
$\bm{I}$ and $\text{diag}(\cdot)$ are the identity matrix and diagonal matrix.
$\text{Tr}(\cdot)$ stands for the matrix trace, and $|| \cdot ||_\text{F}$ is the Frobenius norm.
$\Re(\cdot)$ and $\Im(\cdot)$ denote the real and imaginary parts of complex numbers, and $\textsf{j}=\sqrt{-1}$.
% The transpose and conjugate operators are denoted by $(\cdot )^{\text{T}}$ and $(\cdot )^{*}$, respectively, and the Hermitian transpose is denoted as $(\cdot )^{\text{H}}$.
% while $|\cdot|$ denotes the magnitude operator.
The transpose and conjugate transpose operators are denoted by $(\cdot )^{\text{T}}$ and $(\cdot )^{\text{H}}$, respectively.
$\langle \cdot \rangle$ is used to represent the order statistics, e.g., for $\mathcal{A} = \{a_1, a_2, \cdots , a_n\}$, $\langle \mathcal{A} \rangle = \{a_{\sigma_1}, a_{\sigma_2}, \cdots, a_{\sigma_n}\}$ with $a_{\sigma_1} \leq a_{\sigma_2} \leq \cdots \leq a_{\sigma_n}$.
% $\odot$ stands for element-wise product.
$\mathbb{E}(\cdot)$ represents the mathematical expectation, and $\otimes$ is the Kronecker product.
$\mathcal{CN}\left( \mu, \sigma^2 \right)$ denotes the complex normal distribution with expectation $\mu$ and variance $\sigma^2$.
$C(n,k)$ represents the $k$-combination from the set with $n$ elements.
The notation $~\widetilde{}~$ denotes the angular-delay-domain variables, while the variable with $~\widehat{}~$ is the estimated result by deep learning.

%TDD
%but FDD' problem
%how to tackle this problem is a crucial issue.
%However, the enormous number of antennas imposes huge overhead of channel estimation and feedback
%
%Compress sensing
%angular-delay domain
%R17 codebook
%
%deep learning
%
%the problem of deep learning:
%full CSI?
%huge overhead of downlink channel estimation
%eigenvector

\section{System Model and R17 Type-II codebook}\label{S2}

\subsection{System Model}\label{S2.1}

Consider a FDD MU-MIMO system, where one BS equipped with $N_\text{Tx}$ antennas serves $K$ single-antenna UEs.
Assume that the BS is deployed with dual-polarization antennas in the uniform planar array (UPA), and let $N_\text{h}$ and $N_\text{v}$ represent the numbers of dual-polarization antennas at horizonal and vertical directions, respectively, hence the number of BS antennas $N_\text{Tx}$ can be calculated as $N_\text{Tx} = 2 N_\text{h} N_\text{v}$.
Following the 5G standard \cite{ref35}, we further assume that a cyclic-prefix OFDM system containing $M$ subbands is adopted, where each subband comprises $N_\text{s}$ consecutive resource blocks (RBs) and is regarded the granularity of channel estimation, while each RB consists of 12 consecutive subcarriers.
% The granularity of channel estimation and precoding is defined as the subband that comprises $N_\text{s} \geq 1$ consecutive RBs.
Denoting the CSI vector of the $k$-th UE on the $m$-th subband as $\bm{h}_{k,m} \in \mathbb{C}^{N_\text{Tx} \times 1}$, the corresponding CSI matrix on all subbands $\bm{H}_k \in \mathbb{C}^{N_\text{Tx} \times M}$ can be obtained by concatenating the CSI vectors, i.e., $\bm{H}_k = \left[ \bm{h}_{k,1} ~ \bm{h}_{k,2} ~ ... ~ \bm{h}_{k,M} \right]$.
% classify UL and DL channel symbols?

To generate the CSI matrices, the clustered delay line channel model defined in 3GPP Specification 38.901 \cite{ref42} is adopted.
Specifically, the model contains multiple non-line-of-sight (NLOS) paths and one possible line-of-sight (LOS) path, where each path has its corresponding pathloss, angle-of-arrival (AoA), angle-of-departure (AoD) and delay.
Due to the limited number of scatterers in the propagation environment, the NLOS paths are grouped into several clusters, and the paths in each cluster share similar pathlosses, AoAs, AoDs and delays.
Furthermore, the existing studies have shown that the uplink and downlink channels in FDD systems enjoy the similar power property in angular and delay domains \cite{ref7, ref11, ref12, ref13}.
Consequently, the 38.901 channel model assumes that the uplink and downlink channels for FDD systems have analogous pathlosses, AoAs, AoDs and delays, with slight perturbations depending on the gap of center frequencies.
For convenience, the subscripts UL and DL are adopted in the following to distinguish uplink and downlink channels.

Based on the CSI matrices, linear precoding is assumed to be applied at BS side for simultaneously serving $K$ UEs in the downlink transmissions.
Specifically, let $\bm{v}_{k,m} \in \mathbb{C}^{N_\text{Tx} \times 1}$ denote the precoding vector for the $k$-th UE on the $m$-th subband, and $\bm{V}_{m} = \left[ \bm{v}_{1,m} ~ \bm{v}_{2,m} ~ ... ~ \bm{v}_{K,m} \right] \in \mathbb{C}^{N_\text{Tx} \times K}$ represents the corresponding precoding matrix for all UEs.
Therefore, the BS transmitted signal vector on the $m$-th subband $\bm{x}_m \in \mathbb{C}^{N_\text{Tx} \times 1}$ can be expressed as \cite{ref33}
\begin{align}\label{eq1}
\bm{x}_m = \sum_{k=1}^K  \bm{v}_{k,m} s_{k,m} = \bm{V}_m \bm{s}_m,
\end{align}
where $s_{k,m}$ is the symbol transmitted to the $k$-th UE on the $m$-th subband, and $\bm{s}_{m} = [ s_{1,m} ~ s_{2,m} ~ ...$ $s_{K,m} ]^\text{T} \in \mathbb{C}^{K \times 1}$ represents the symbol vector. The precoding matrix $\bm{V}_m$ satisfies the constraint of the total power $P_\text{Tx}$, i.e., $\sum_{m=1}^{M}\text{Tr}(\bm{V}_m \bm{V}_m^\text{H}) \leq P_\text{Tx}$, and the symbol vector $\bm{s}_m$ is normalized as $\mathbb{E}\left(\bm{s}_m \bm{s}_m^\text{H}\right) = \bm{I}$.
Accordingly, the received signal of the $k$-th UE on the $m$-th subband $y_{k,m}$ can be obtained as
\begin{align}\label{eq2}
 y_{k,m} = \bm{h}_{k,m,\text{DL}}^\text{H} \bm{v}_{k,m} s_{k,m} + \sum_{j \neq k} \bm{h}_{k,m,\text{DL}}^\text{H} \bm{v}_{j,m} s_{j,m} + n_k,
\end{align}
where $n_k$ is the additive white Gaussian noise (AWGN) with power $\sigma^2$, i.e., $n_k \sim \mathcal{CN} \big(0,\sigma^2 \big)$.
Based on the received signal model in (\ref{eq2}), the achievable rate of the $k$-th UE on the $m$-th subband $R_{k,m}$ can be calculated as
\begin{align}\label{eq3}
 R_{k,m} = \log_2\left( 1 + \frac{|\bm{h}_{k,m,\text{DL}}^\text{H} \bm{v}_{k,m}|^2}{\sum_{j \neq k} |\bm{h}_{k,m,\text{DL}}^\text{H} \bm{v}_{j,m}|^2 + \sigma^2} \right),
\end{align}
and the average sum rate of all $K$ UEs on $M$ subbands $R_\text{avg}$ can be expressed as
\begin{align}\label{eq4}
R_\text{avg} = \frac{1}{M} \sum_{m=1}^{M} \sum_{k=1}^{K} R_{k,m}.
\end{align}
% More details can be found in 3GPP specification 38.901 [].
% zenith azimuth

% antenna ports in the horizontal and vertical dimensions in
% one polarization direction
% consisting of one low-frequency BS and $J$ mmWave BSs, where each user accesses the low-frequency BS and one of the $J$ mmWave BSs, as shown in Fig.~\ref{Fig1_b}.
% For simplicity, we consider single UE equipped with one low-frequency antenna and one mmWave antenna \cite{Ref:DL1}, but our scheme can be straightforwardly extended to the multiuser scenario with multiple UE antennas.
% Further assume the two-dimensional system model, namely, BSs are equipped with uniform linear arrays (ULAs), where only azimuth angles are considered.

\subsection{R17 Type-II Codebook}\label{S2.1}

To improve the sum rate performance, accurately acquiring the downlink CSI of each UE is of vital necessity.
However, directly feeding back the whole CSI matrix to BS in FDD systems could impose unacceptably huge overhead, thus how to reduce the overhead of CSI feedback with guaranteeing high CSI accuracy arises as an important issue.
To this end, 3GPP standardized the Type-II codebook to exploit the channel sparse characteristics for supporting low-overhead CSI feedback.
First of all, only the CSI eigenvector is required to be fed back in the Type-II codebook.
In Release 15 (R15), based on the angular-domain sparsity, the Type-II codebook selects the subset of dominant beams with corresponding linear combination coefficients to compress the CSI eigenvector of each subband separately \cite{ref38}.
In R16, the delay-domain sparsity is further harnessed in the Type-II codebook, where the CSI eigenvectors on multiple subbands are transformed into the delay domain for selecting a small fraction of the strongest angular-delay-domain ports for CSI feedback \cite{ref36}.
In spite of effectively reducing the feedback overhead, the Type-II codebooks in R15 and R16 need to measure the whole CSI eigenvectors for extracting the sparse features, which still leads to considerable downlink pilot overhead for CSI measurement.
To address this issue, the R17 Type-II codebook further utilizes the angular-delay-domain partial reciprocity between uplink and downlink channels to reduce the downlink pilot overhead \cite{ref35}, which is the main scenario that we concentrate on in this paper.

Specifically, the R17 Type-II codebook comprises four steps, i.e., uplink angular-delay-domain port selection, downlink port coefficient measurement, uplink port coefficient feedback and downlink CSI reconstruction. For simplicity, we assume single-antenna UEs in the system model, so that the CSI vector on each subband $\bm{h}_{k,m}$ itself is the CSI eigenvector and the SVD operation is not required.

\textbf{Uplink angular-delay-domain port selection}: To support reliable uplink transmissions, the uplink CSI is periodically estimated at BS side, thus the BS possesses the uplink CSI matrices of each UE. To fully capture the channel sparsity, the uplink CSI matrix of the $k$-th UE $\bm{H}_{k,\text{UL}} \in \mathbb{C}^{N_\text{Tx} \times M}$ is firstly transformed into the angular-delay domain as below
\begin{align}\label{eq5}
\bm{H}_{k,\text{UL}} = \bm{W}_\text{A} \widetilde{\bm{H}}_{k,\text{UL}} \bm{W}_\text{D}^\text{H},
\end{align}
where $\widetilde{\bm{H}}_{k,\text{UL}} \in \mathbb{C}^{N_\text{Tx} \times M}$ is the corresponding angular-delay-domain CSI matrix\footnote{The R17 Type-II codebook considers that the uplink CSI for port selection has the same number of subbands as the downlink CSI to be estimated, which can be acquired by selecting part of the uplink CSI from all uplink subbands.}.
In the left, the angular-domain transform matrix $\bm{W}_\text{A} \in \mathbb{C}^{ N_\text{Tx} \times N_\text{Tx} }$ consists of two identical matrices $\bm{D} \in \mathbb{C}^{ N_\text{h} N_\text{v} \times N_\text{h} N_\text{v} }$ on the diagonal corresponding to dual polarization directions, i.e., $\bm{W}_\text{A} = \text{diag}\left( \bm{D}, \bm{D} \right) $.
% To better extract the sparse structure, the oversampled orthogonal discrete Fourier transform (DFT) bases are further considered in $\bm{W}_\text{A}$.
To fully extract the sparse structure, the oversampled orthogonal discrete Fourier transform (DFT) bases are considered in $\bm{D}$.
Specifically, denote $O_\text{h}$ and $O_\text{v}$ as the oversampling factors of the horizonal and vertical directions in the UPA, so that the horizonal DFT bases $\bm{w}^\text{(h)}_{\theta_1}, \theta_1 = 0, 1, ..., O_\text{h} N_\text{h} - 1$ and the vertical DFT bases $\bm{w}^\text{(v)}_{\theta_2}, \theta_2 = 0, 1, ..., O_\text{v} N_\text{v} - 1$ can be separately represented as
\begin{align}\label{eq67}
\bm{w}^\text{(h)}_{\theta_1} = \left[ 1 ~ e^\textsf{j} \frac{2 \pi {\theta_1}}{O_\text{h} N_\text{h}} ~ ... ~ e^\textsf{j} \frac{2 \pi {\theta_1} (N_\text{h} - 1)}{O_\text{h} N_\text{h}} \right]^\text{T}, \\
\bm{w}^\text{(v)}_{\theta_2} = \left[ 1 ~ e^\textsf{j} \frac{2 \pi {\theta_2}}{O_\text{v} N_\text{v}} ~ ... ~ e^\textsf{j} \frac{2 \pi {\theta_2} (N_\text{v} - 1)}{O_\text{v} N_\text{v}} \right]^\text{T}.
\end{align}
For the whole UPA, the DFT basis can be obtained as $\bm{w}^{\text{(h,v)}}_{\theta_1,\theta_2} = \bm{w}^\text{(h)}_{\theta_1} \otimes \bm{w}^\text{(v)}_{\theta_2} \in \mathbb{C}^{ N_\text{h} N_\text{v} \times 1 }$, which forms a column in $\bm{D}$. By carefully choosing the orthogonal combination of angular-domain bases with better sparsity, a small number of bases could capture larger channel power.
Similarly, the delay-domain transform matrix $\bm{W}_\text{D} \in \mathbb{C}^{ M \times M }$ in the right comprises $M$ orthogonal DFT bases.

Once transformed into the angular-delay domain, the BS can select $P$ angular-delay-domain ports with the highest power for downlink channel measurement, where each port corresponds to a specific angular-delay-domain basis.

\textbf{Downlink port coefficient measurement}: Since the selected $P$ angular-delay-domain ports occupy a vast majority of channel power, the BS can only transmit the downlink pilots on these ports to acquire a sufficiently accurate CSI estimate. Specifically, assume that the $p$-th port for the $k$-th UE adopts the $p^\text{(A)}_{k}$-th column in $\bm{W}_\text{A}$, i.e., $\bm{w}_{\text{A},p^\text{(A)}_{k}} = {\bm{W}_\text{A}}{[:, p^\text{(A)}_{k}]}$, as the angular-domain basis, and adopts the $p^\text{(D)}_{k}$-th column in $\bm{W}_\text{D}$, i.e., $\bm{w}_{\text{D},p^\text{(D)}_{k}} = {\bm{W}_\text{D}}{[:,p^\text{(D)}_{k}]}$, as the delay-domain basis.
%Consequently, the corresponding downlink angular-delay-domain pilot matrix $\bm{\Phi}_{k,p} \in \mathbb{C}^{ N_\text{Tx} \times M }$ can be calculated as $\bm{\Phi}_{k,p} = \sqrt{P_\text{Tx}} \bm{w}_{\text{A},p^\text{(A)}_{k}} \bm{w}_{\text{D},p^\text{(D)}_{k}}^\text{H}$, and its port coefficient $c_{k,p}$ is obtained by downlink pilot measurement as $c_{k,p} = \text{Tr}\left( \bm{H}_k \bm{\Phi}_{k,p}^\text{H} \right)$.
Consequently, the corresponding normalized downlink precoding matrix for port measurement $\bm{\Phi}_{k,p} \in \mathbb{C}^{ N_\text{Tx} \times M }$ can be expressed as $\bm{\Phi}_{k,p} = \bm{w}_{\text{A},p^\text{(A)}_{k}} \bm{w}_{\text{D},p^\text{(D)}_{k}}^\text{H}$, and its port coefficient $c_{k,p}$ is calculated as $c_{k,p} = \text{Tr}\left( \bm{H}_{k,\text{DL}} \bm{\Phi}_{k,p}^\text{H} \right)$.
%Hence the original channel matrix can be reconstructed as
%\begin{align}\label{eq7}
%\overline{\bm{H}}_{k} = \sum_{p=1}^{P} c_{k,p} \bm{w}_{\text{A},p_\text{A}}^\text{T} \bm{w}_{\text{D},p_\text{D}}.
%\end{align}
%\begin{align}\label{eq2}
%c_{k,p} = \text{Tr}\left( \bm{\Phi}_{k,p}^\text{H} \bm{H}_k \right),
%\end{align}

\textbf{Uplink port coefficient feedback}: In order to reduce the overhead of feeding back the port coefficients, the R17 Type-II codebook applies a tailored quantization method that comprises two stages. In the first stage, the $Q_\text{w}$-bit wideband amplitude quantization is introduced to indicate the amplitude ratio of the polarization direction with smaller maximum amplitude to that with larger maximum amplitude, which avoids that the port coefficients in the weaker polarization direction are buried by quantization errors.
Then, the amplitude and phase of each narrowband port coefficient are separately compressed by the $Q_\text{n,a}$-bit logarithmic quantization and $Q_\text{n,p}$-bit uniform quantization in the second stage.
After quantization, the quantized port coefficients $\bar{c}_{k,p}, p = 1,2,..., P$ are fed back from the $k$-th UE to BS.

\textbf{Downlink CSI reconstruction}: Finally, the reconstructed downlink CSI matrix of the $k$-th UE $\bm{H}_{k,\text{DL(TypeII)}}$ can be expressed as below
\begin{align}\label{eq8}
\bm{H}_{k,\text{DL(TypeII)}} = \sum_{p=1}^{P} \bar{c}_{k,p} \bm{w}_{\text{A},p^\text{(A)}_{k}} \bm{w}_{\text{D},p^\text{(D)}_{k}}^\text{H}.
\end{align}
Based on the reconstructed CSI matrices at BS side, various precoding methods can be applied to reduce the inter-UE interference for attaining high sum rate performance.
In summary, the R17 Type-II codebook only requires $P \ll N_\text{Tx} M$ angular-delay-domain ports for downlink CSI measurement as well as at most $\left[Q_\text{w} + P(Q_\text{n,a} + Q_\text{n,p}) \right]$ bits for uplink CSI feedback, which effectively relieves the heavy burden of BS downlink CSI acquisition in FDD systems.

%Since the selected angular-dolay ports occupy a vast majority of channel power, BS can only transmit the downlink pilots on these ports to acquire a sufficiently accurate downlink channel estimate. Specifically, assume the $p$-th port adopts the $p_A$-th column in $\bm{W}_\text{A}$, i.e., $\bm{w}_{\text{A},p_\text{A}} = {\bm{W}_\text{A}}_{\left[:, p_\text{A}\right]}$, as the angular basis, and adopt the $p_D$-th row in $\bm{W}_\text{D}$, i.e., $\bm{w}_{\text{D},p_\text{D}} = {\bm{W}_\text{D}}_{\left[p_\text{D}, :\right]}$, as the delay-domain basis.
%$k$ index?
%Consequently, the corresponding downlink angular-delay-domain pilot matrix for the $k$-th UE $\bm{\Phi}_{k,p} \in \mathbb{C}^{ N_\text{Tx} \times M }$ can be calculated as $\bm{\Phi}_{k,p} = \bm{w}_{\text{A},p_\text{A}}^\text{T} \bm{w}_{\text{D},p_\text{D}}$, and the corresponding port coefficient $c_{k,p}$ is obtained as $c_{k,p} = \text{Tr}\left( \bm{\Phi}_{k,p}^\text{H} \bm{H}_k \right)$.
%Hence the original channel matrix can be reconstructed as
%\begin{align}\label{eq2}
%\overline{\bm{H}}_{k} = \sum_{p=1}^{P} c_{k,p} \bm{w}_{\text{A},p_\text{A}}^\text{T} \bm{w}_{\text{D},p_\text{D}}.
%\end{align}
%pilot design

% = \text{diag}\left( \bm{D}, \bm{D} \right) \in \{ \}$

\section{Deep Learning Based Angular-Delay-Domain Port Selection}\label{S3}

\subsection{Motivation and Problem Formulation}\label{S3.1}

Although the R17 Type-II codebook exhibits the outstanding performance in reducing the overhead of downlink CSI measurement and feedback in FDD systems, its accuracy of CSI reconstruction heavily relies on the selected angular-delay-domain ports based on the uplink CSI.
In other words, the reconstructed CSI would be more accurate if the selected ports can capture a larger proportion of CSI power.
However, since the transmit power at UE side is quite limited (e.g., $0.2~\text{W}$ as a typical value \cite{ref45}), the measured uplink CSI usually suffers from a relatively low SNR, which incurs the inaccuracy of port selection and degrades the performance of the Type-II codebook.

Fortunately, owing to the limited number of scatterers in the propagation environment around BS and UE, most of the paths in the channel come from a few angular and delay ranges in general.
This characteristic makes the CSI matrix possess the sparse structure in the angular-delay domain \cite{ref6, ref7}, which brings the correlations between the coefficients of different ports that can be used for enhancing the accuracy of port selection.
For instance, the DFT bases adopted in the Type-II codebook may cause the power leakage effect, i.e., when the direction of the angular-delay-domain port is not aligned with the physical path, a determined amount of path power would leak to the neighboring ports according to the misalignment degree \cite{ref46, ref47}.
Therefore, the port correlations resulting from this effect can be harnessed to redress wrong port selections in the relatively low-SNR scenarios.
% For instance, the DFT bases adopted in the Type-II codebook may cause the power leakage effect, i.e., when the direction of the angular-delay-domain port is not aligned with the physical path, a determined amount of path power according to the misalignment degree would leak to the neighboring ports \cite{ref46, ref47}.
%Consequently, it is more likely to select the angular-delay-domain ports around the dominant port.
Nevertheless, the sparse structure in the multipath environment is fairly complicated and hard to be accurately extracted by the conventional methods.
Inspired by the strong adaptive fitting capabilities, deep learning is adopted to capture the sparse features for accurately selecting the optimal angular-delay-domain ports.

Specifically, let $\mathcal{P}_{k}$ denote the index set of selected angular-delay-domain ports for the $k$-th UE.
Considering that the port selection process is the same for all UEs, the subscript of UE index $k$ is omitted in the following of this section.
Accordingly, the port selection problem can be written as
\begin{align}\label{eq9}
\mathcal{P} = f\left(\bm{H}_\text{UL} \right),
\end{align}
where $f(\cdot)$ represents the port selection function.
For the selection of single optimal port, the traditional studies usually formulate this problem as a multi-classification task, where each possible selection is viewed as a class \cite{ref48, ref49, ref50}.
However, the output in (\ref{eq9}) is a set that contains $C \left( N_\text{Tx} M, P \right)$ possible combinations, hence it is not feasible to establish a multi-classification task for solving (\ref{eq9}).
% To address this problem, we observe that (\ref{eq9}) can be decomposed into $N_\text{Tx} M$ subproblems, where the $n$-th subproblem judges whether the $n$-th port is selected according to its own label.
To address this problem, we observe that (\ref{eq9}) can be decomposed into $N_\text{Tx} M$ subproblems, where the $n$-th subproblem judges whether the $n$-th port is selected.
Consequently, (\ref{eq9}) is transformed into a multi-label classification task with $N_\text{Tx} M$ labels as below
\begin{align}
& I_n = f_n\left(\bm{H}_\text{UL} \right), n \in \left\{ 1, 2, ..., N_\text{Tx} M \right\}, \label{eq10_basic} \\
& \text{s.t.}~\sum_{n=1}^{N_\text{Tx} M} I_n = P, \tag{\ref{eq10_basic}{a}} \label{eq10a}
\end{align}
where the $n$-th label $I_n$ satisfies $I_n = 1$ if the $n$-th port is selected and otherwise $I_n = 0$, while $f_n(\cdot)$ denotes the corresponding selection function.
By tackling the binary-classification tasks for each label under the constraint that the total number of positive classifications equals to $P$, the selected port set $\mathcal{P}$ can be acquired as the indices with positive classifications.
%\begin{align}
%& I_n = f_n\left(\bm{H}_\text{UL} \right), n = 1, 2, ..., N_\text{Tx} M, \label{eq10_basic} \\
%& \text{s.t.}~\sum_{n=1}^{N_\text{Tx} M} I_n = P, \tag{\ref{eq10_basic}{a}} \label{eq10a}
%\end{align}
% where $I_n$ satisfies $I_n = 1$ if the $n$-th port is selected and otherwise $I_n = 0$.

\subsection{Proposed Model Design}\label{S3.2}

Because the sparse structure of CSI brings the correlations between the selection of different ports, a unified deep learning model is constructed to extract the CSI features for all $N_\text{Tx} M$ subproblems, which exhibits $N_\text{Tx} M$ binary outputs as the classification result.
In consistent with many existing studies on deep learning based CSI feedback \cite{ref16,ref17,ref18,ref19,ref20}, the convolutional neural network (CNN) is adopted as the backbone architecture for port selection. Specifically, our model design consists of three components, the preprocessing module, the convolutional module and the output module, as shown in Fig.~1.

\textbf{1) Preprocessing Module}: To better capture the sparse features, the uplink CSI matrix $\bm{H}_\text{UL}$ is firstly transformed to the angular-delay-domain CSI $\widetilde{\bm{H}}_\text{UL}$ as the model input.
Next, considering the small amplitude of $\widetilde{\bm{H}}_\text{UL}$, an input normalization operation is applied to normalize the maximum amplitude of the elements in $\widetilde{\bm{H}}_\text{UL}$ into $1$.
Besides, the normalized CSI $\widetilde{\bm{H}}_\text{UL}^\text{N}$ is divided into two real-valued feature channels corresponding to its real and imaginary parts $\left\{ \Re \left( \widetilde{\bm{H}}_\text{UL}^\text{N} \right), \Im \left( \widetilde{\bm{H}}_\text{UL}^\text{N} \right) \right\}$, which are fed into the following convolutional module.

\begin{figure}[tp!]
% \vspace{-6mm}
\begin{center}
\includegraphics[width=1.0\textwidth]{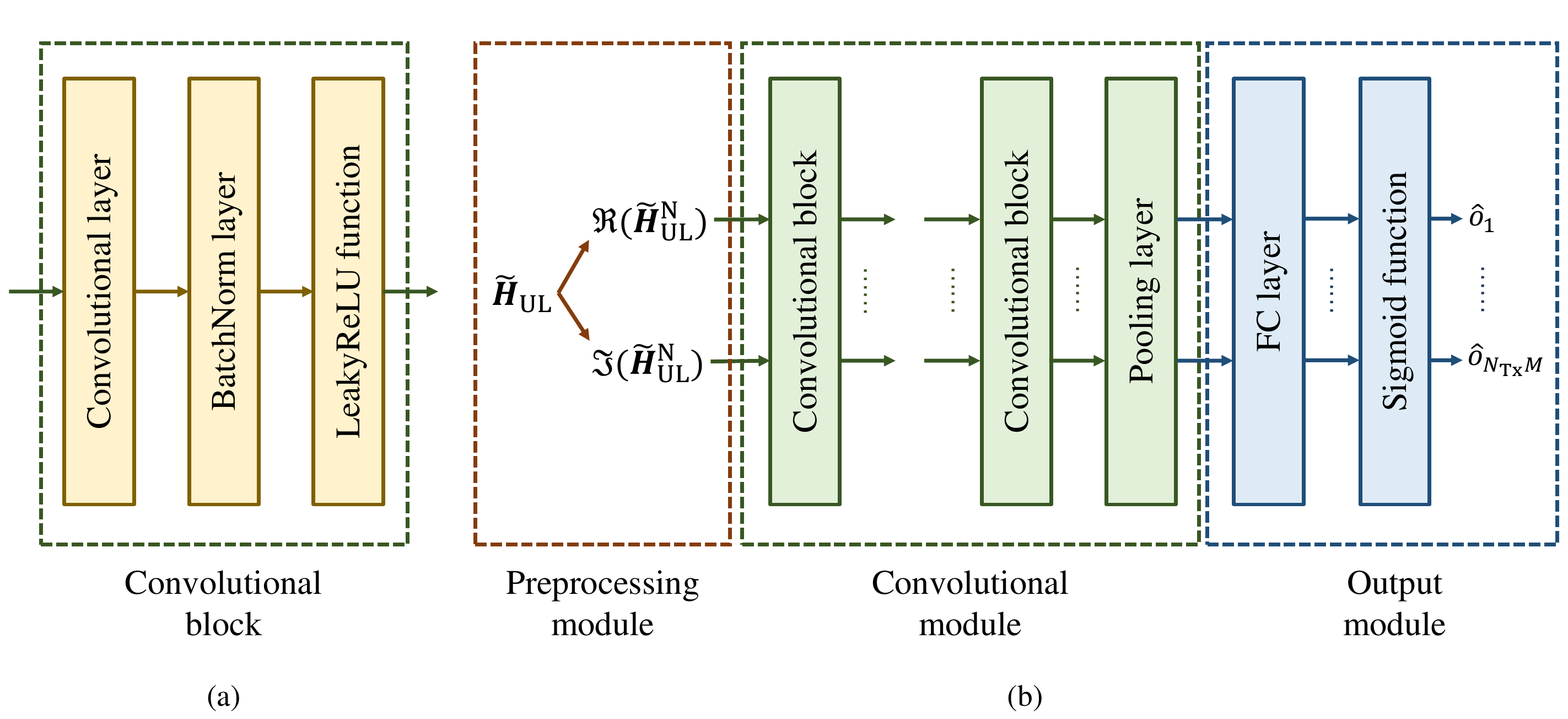}
\end{center}
\vspace{-8mm}
\caption{Illustrations of (a) convolutional block and (b) proposed deep learning model for port selection, which consists of preprocessing module, convolutional module and output module.}
\label{fig1} % Fig.8
\vspace{-6mm}
\end{figure}

\textbf{2) Convolutional module}: Multiple convolutional blocks are adopted to learn the sparse features, where each block is made up of a convolutional layer, a batch normalization (BatchNorm) layer and a LeakyReLU activation function in order, as depicted in Fig.~1(a).
Concretely, the convolutional layer adaptively extracts the hidden features based on its learnable convolutional kernels, where the padding operation is utilized to expand the input edges to avoid obliterating useful information on the boundaries.
Then, the BatchNorm layer normalizes the extracted features to a standard distribution with mean $0$ and variance $1$ for stabilizing model convergence \cite{ref51}, and the LeakyReLU activation function is applied to provide the nonlinear fitting capability.
After the ultimate convolutional block, a pooling layer is used to downsample the output matrix in each feature channel to a scalar.

In order to enhance the accuracy of port selection, the circular padding is proposed to be exploited in the convolutional layers \cite{ref52}.
Specifically, the two opposite edges of the angular-delay-domain CSI are actually `adjacent' because of the cyclic property of DFT bases.
However, the broadly applied zero padding ignores this property and directly adds zeros at the edges, which buries their correlations and results in inadequate learning of the sparse structure.
By contrast, the circular padding adopts the opposite edge as the padding contents, so that the correlations between the edges can be fully reserved.

%Besides, it is observed the widely applied zero-padding in the convolution layers could lead to accuracy loss in the port selection.
%Specifically, the two opposite edges of the angular-delay domain CSI are actually `adjacent' because of the cyclic property of DFT bases, whereas the zero-padding ignores this property and directly adds zeros at the edges, which buries their correlations and results in insufficient learning of the sparse structure.
%To tackle this problem, we propose to exploit the circular-padding in the convolution layers \cite{ref51}, where the features at the opposite edge are adopted as the padding contents, so that the correlations between the edges can be fully reserved.

\textbf{3) Output module}:
In order to carry out the multi-label classification task, a fully-connected (FC) layer is introduced after the pooling layer to implement the transformation from the extracted features to the angular-delay-domain ports, followed by a sigmoid activation function to normalize the range of each output into $(0, 1)$, which can be expressed as
\begin{equation}\label{eq11}
\widehat{o}_{n} = \text{sigmoid}\big(\bm{u}^{\text{T}}_n \bm{z} + b_n\big),n \in \{1,2,...,N_\text{Tx} M\},
\end{equation}
where $\bm{z} \in \mathbb{R}^{Z \times 1}$ is the output feature vector of the pooling layer with length $Z$, while $\bm{u}_n \in \mathbb{R}^{Z \times 1}$ and $b_n$ denote the weights and bias of the FC layer for the $n$-th output $\widehat{o}_n$.
The output vector $\widehat{\bm{o}} = \left[ \widehat{o}_{1} ~ \widehat{o}_{2} ~ ... ~ \widehat{o}_{N_\text{Tx} M} \right]^\text{T}$ illustrates the priority of selecting the angular-delay-domain ports, i.e., the $n$-th port is preferred to be selected if it has higher $\widehat{o}_{n}$.
Therefore, to meet the constraint of total selected port number, the $P$ angular-delay-domain ports with highest outputs are selected as below
\begin{align} % eqs.12-13
 & \left\{ \widehat{o}_{\sigma_1}, \widehat{o}_{\sigma_2}, ..., \widehat{o}_{\sigma_{N_\text{Tx} M}} \right\} = \left\langle \left\{ \widehat{o}_{1}, \widehat{o}_{2}, ..., \widehat{o}_{N_\text{Tx} M} \right\} \right\rangle, \label{eq12} \\
 & \mathcal{P} = \left\{ \sigma_{N_\text{Tx} M - P + 1}, \sigma_{N_\text{Tx} M - P + 2}, ..., \sigma_{N_\text{Tx} M} \right\}. \label{eq13}
\end{align}

% Based on the model optimization, the output of the selected port
% To meet the constraint of total selected port number, the

% To predict the optimal narrow beam from all the candidate narrow beams, the fully-connected (FC) layer is introduced after the pooling layer to implement the transformation from the extracted features to the candidate narrow beams, followed by a softmax activation layer for normalizing the outputs into probabilities, which can be written as
% where $\widehat{p}_{{\rm LF},n}$ is the predicted probability that the $n$-th candidate narrow beam is the optimal one, and $\bm{v}$ is the output vector of the pooling layer, while $\bm{u}_n$ and $b_n$ are the weight vector and bias of the FC layer corresponding to the $n$-th output. For convenience, we define the probability vector of the narrow beams predicted from the low-frequency CSI as $\widehat{\bm{p}}_{\rm LF} = \left[ \widehat{p}_{{\rm LF},1} ~ \widehat{p}_{{\rm LF},2} \cdots \widehat{p}_{{\rm LF},J N} \right]^{\rm T}$.

\subsection{Proposed Model Training with Focal Loss}\label{S3.3}

To optimize the parameters of our proposed deep learning model, the binary cross entropy (BCE) loss is utilized for each angular-delay-domain port selection, and the loss of the whole model can be written as
\begin{align}\label{eq14}
 \text{loss}_\text{BCE} = - \frac{1}{N_\text{Tx}M}\sum\limits_{n=1}^{N_\text{Tx}M} \big[ I_n \log\widehat{o}_n + (1 - I_n) \log\left(1 - \widehat{o}_n\right) \big].
\end{align}
Recall that $I_n$ represents the label for the selection of the $n$-th port.

Nevertheless, directly using (\ref{eq14}) to train the model may not achieve an adequately accurate port selection for two reasons.
Firstly, since the number of selected ports $P$ is generally much smaller than that of all ports $N_\text{Tx} M$, the optimization process would mainly pay attention to the discarded ports and underrate the selected ports with more importance.
Secondly, it is obviously more difficult to accurately select the port with smaller power than that with larger power from all the $P$ optimal ports, whereas the loss function in (\ref{eq14}) provides the equal weight for both simple and difficult ports, which could lead to insufficient model training.
To address these two issues, the focal loss \cite{ref53} is introduced to balance the training samples for enhancing the selection accuracy.
The focal loss was originally proposed to improve the performance of dense object detection with a serious imbalance problem, which is very suitable to be integrated with our port selection task.
Specifically, the focal loss firstly adds the weights to balance the losses of positive class (i.e., the selected ports) and negative class (i.e., the discarded ports), where the ratio of loss weights can be empirically set as the reciprocal of the ratio of sample numbers.
Then, the positive and negative labels are separately scaled by the modulating factors $\left\{(1 - \widehat{o}_n)^\gamma, \widehat{o}_n^\gamma\right\}$ with a tunable focusing
parameter $\gamma \geq 0$, which could aid the loss function to concentrate more on the difficult ports.
For example, if the output $\widehat{o}_n$ of a positive port is very small, the focal loss views it as a difficult port and thus the modulating factor $(1 - \widehat{o}_n)^\gamma$ would be close to $1$.
By contrast, the positive port with the output $\widehat{o}_n$ near $1$ is regarded as a simple port and would have a small modulating factor.
In summary, the BCE loss of our model improved by the focal loss (FL) is expressed as
\begin{align}\label{eq15}
 \text{loss}_\text{BCE,FL} = & - \frac{1}{N_\text{Tx}M}\sum\limits_{n=1}^{N_\text{Tx}M} \Big[ \frac{N_\text{Tx}M - P}{N_\text{Tx}M} (1 - \widehat{o}_n)^\gamma I_n \log\widehat{o}_n + \nonumber \\
  & \frac{P}{N_\text{Tx}M} \widehat{o}_n^\gamma (1 - I_n) \log\left(1 - \widehat{o}_n\right) \Big].
\end{align}

The practical deployment of our proposed deep learning model consists of two stages, training and predicting.
At the training stage, training data are collected to optimize the model, where each sample comprises the noisy uplink CSI matrix as the input, together with the indices of the $P$ optimal angular-delay-domain ports as the classification label.
The labels can be obtained based on the high-SNR angular-delay-domain CSI matrices estimated by long pilot sequences.
Throughout this stage, the back propagation algorithm is harnessed to train the model parameters according to the loss function in (\ref{eq15}).
Once the model is well trained with adequate data, it switches to the predicting stage.
At this stage, the BS adopts the estimated uplink CSI matrix to select the strongest angular-delay-domain ports based on the model, which could achieve higher selection accuracy and improve the performance of the Type-II codebook.

% the qualities of all the candidate mmWave wide beams are predicted based on the low-frequency CSI, and only part of the predicted high-quality mmWave wide beams are measured. Then the low-frequency CSI and the received signal vector of the measured mmWave wide beams are adopted to jointly predict the optimal mmWave narrow beam. Therefore, BS and UE only measure the low-frequency CSI and a few mmWave wide beams, and thus the huge overhead of both mmWave BS selection and beam training is significantly reduced.

\section{Deep Learning Based Angular-Delay-Domain Channel Reconstruction}\label{S4}

\subsection{Motivation and Problem Formulation}\label{S4.1}

In the previous section, we focus on addressing the inaccuracy issue in the angular-delay-domain port selection incurred by relatively low SNRs of uplink CSI.
Assuming the perfect port selection, the quantization errors of feeding back the port coefficients of downlink CSI become another performance bottleneck of R17 Type-II codebook.
% Most of the existing deep learning based CSI feedback studies utilized the AE architecture to improve the performance, where the deep learning based encoder deployed at UE side firstly compresses the full CSI matrices \cite{ref16}--\cite{ref29} or CSI eigenvectors \cite{ref36,ref37,ref_add1,ref38} for feedback, while the decoder at BS side reconstructs the CSI by deep learning.
To tackle this problem, most of the existing deep learning based works utilize the AE to implement the feedback \cite{ref16,ref17,ref18,ref19,ref20,ref21,ref22,ref23,ref24,ref25,ref26,ref27,ref28,ref36,ref37,ref_add1,ref38}, where the encoder deployed at UE side firstly compresses the full CSI matrices or CSI eigenvectors for feedback, while the decoder at BS side correspondingly reconstructs the CSI.
% Most of the existing deep learning based CSI feedback studies utilized the AE architecture to improve the performance, where the deep learning based encoder deployed at UE side firstly compresses the CSI for feedback, while the decoder at BS side reconstructs the CSI by deep learning.
Relying on the determined CSI structure, these traditional methods can sufficiently exploit the sparse features reflected by the correlations between different CSI elements to improve the feedback performance.
% Nevertheless, this advantage cannot be well leveraged for the R17 Type-II codebook, since the UE does not know the corresponding angular-delay-domain ports of the measured CSI coefficients.
Nevertheless, this advantage cannot be well leveraged for the R17 Type-II codebook, because the selected angular-delay-domain ports for the UEs at diverse locations are quite different in practical scenarios.
% In the practical scenarios, the selected ports for the UE at diverse locations are quite different, so that the coefficient vector does not have the determined correlations in the eyes of the encoder at UE side.
Considering that the UE does not know the corresponding angular-delay-domain bases of the measured port coefficients, the coefficient vector for feedback does not have the determined intrinsic correlations in the views of the encoder at UE side.
%Briefly speaking, this coefficient vector loses its sparse structure.
More seriously, the existing theoretical analysis demonstrated that deep learning suffers from the low-rank bias \cite{ref40, ref41}, i.e., deep learning tends to extract the low-rank sparse features and is not good at handling irregular non-sparse data, which may degrade the performance of compressing the coefficient vector of R17 Type-II codebook based on deep learning.
In addition, the considerable calculational overhead of deep learning based CSI compression at UE side is a vital concern for practical deployments \cite{ref54}.

To solve the problems above, we propose to apply the standard quantization process in the R17 Type-II codebook to compress the coefficient vector of selected ports for feedback, and focus on the CSI reconstruction at BS side.
Because the BS is aware of the angular-delay-domain basis for each feedback coefficient, it can place all the coefficients on their corresponding angular-delay-domain ports, which forms the reconstructed CSI by the Type-II codebook according to (\ref{eq8}).
It is clear that the sparse structure in this reconstructed CSI is fully reserved, thus deep learning can be exploited to extract the sparse features for enhancing the reconstruction accuracy.
Furthermore, the coefficients in the discarded ports can also be reconstructed according to the sparse structure, since the partial power of these ports may come from the leakage of the selected dominant ports.

To be specific, recall that $\bm{H}_{k,\text{DL(TypeII)}} \in \mathbb{C}^{N_\text{Tx} \times M}$ denotes the reconstructed downlink CSI for the $k$-th UE based on the feedback of the Type-II codebook, while let $\bm{H}_{k,\text{DL(p)}} \in \mathbb{C}^{N_\text{Tx} \times M}$ represent the corresponding perfect reconstructed downlink CSI, which contains the coefficients of all $N_\text{Tx} M$ ports without quantization errors.
Consequently, the deep learning based CSI reconstruction problem can be formulated as a regression task as below
\begin{align}\label{eq16}
\bm{H}_{k,\text{DL(p)}} = g\left( \bm{H}_{k,\text{DL(TypeII)}} \right),
\end{align}
where $g(\cdot)$ denotes the regression function.
To better capture the correlations of port coefficients, we transform the reconstruction problem (\ref{eq16}) to the angular-delay domain, i.e., $ \widetilde{\bm{H}}_{k,\text{DL(p)}} = g_0 \left( \widetilde{\bm{H}}_{k,\text{DL(TypeII)}} \right) $ with its corresponding regression function $g_0 (\cdot)$.
% Next, we will focus on designing the deep learning model to accurately fit $g(\cdot)$.

\subsection{Proposed Model Design with Weighted Shortcut}\label{S4.2}

\begin{figure}[bp!]
\vspace{-5mm}
\begin{center}
\includegraphics[width=.8\textwidth]{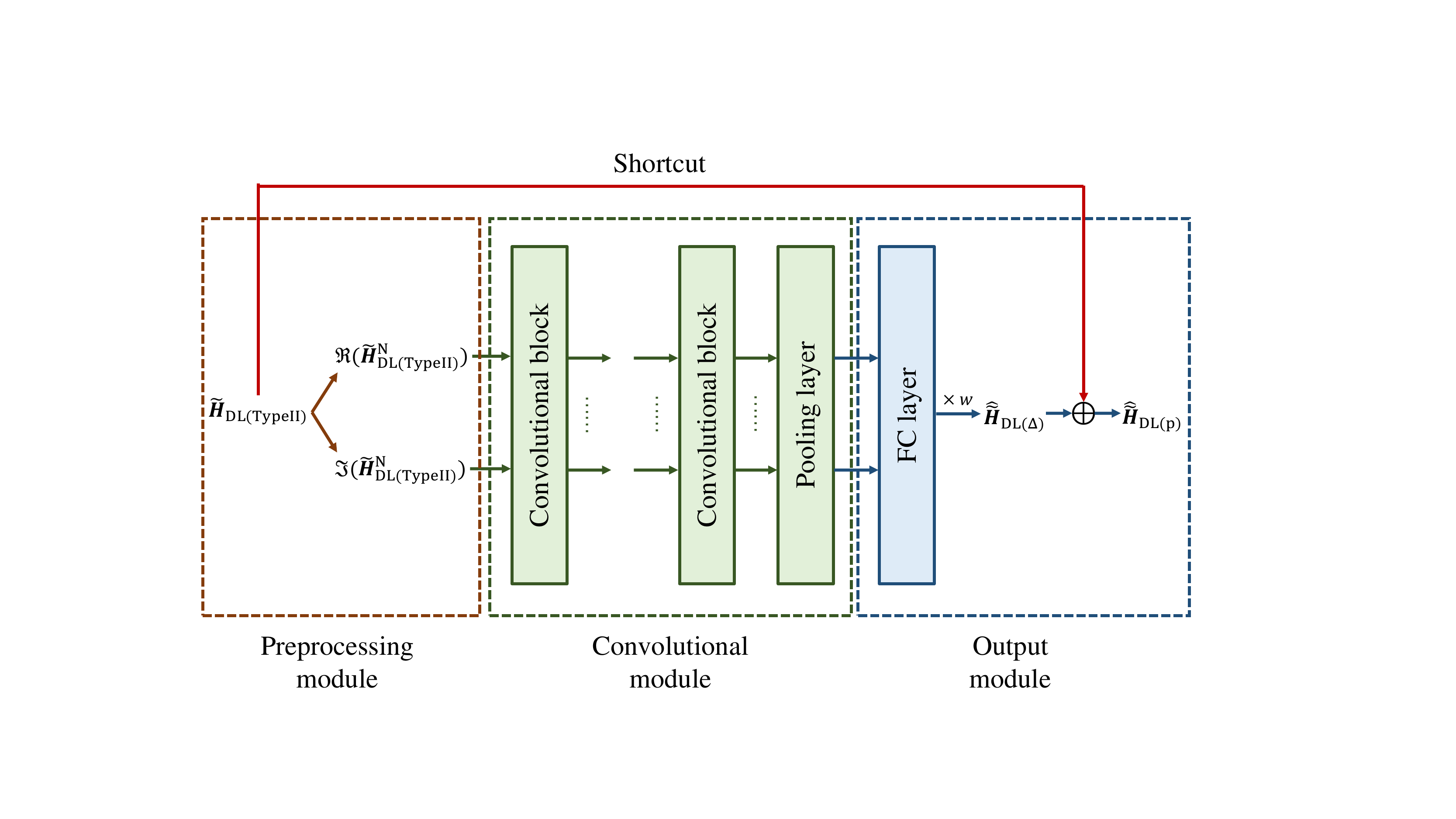}
\end{center}
\vspace{-8mm}
\caption{Illustration of proposed deep learning model with weighted shortcut for CSI reconstruction, where the subscript of UE index $k$ is omitted.}
\label{fig2} % Fig.2
%\vspace{-5mm}
\end{figure}

Similar to Subsection \ref{S3.2}, CNN is adopted as the backbone architecture to implement the CSI reconstruction, as depicted in Fig.~\ref{fig2}. The key difference lies in the output module, so we only detail this module in the subsection for brevity.

Considering that the selected ports usually occupy the vast majority of CSI power and the quantized port coefficients are close to their original values, the reconstructed angular-delay-domain CSI based on the Type-II codebook $\widetilde{\bm{H}}_{k,\text{DL(TypeII)}}$ has high similarity to its perfect counterpart $\widetilde{\bm{H}}_{k,\text{DL(p)}}$.
Therefore, to alleviate the burden of deep learning, we propose to regress the CSI difference between $\widetilde{\bm{H}}_{k,\text{DL(TypeII)}}$ and $\widetilde{\bm{H}}_{k,\text{DL(p)}}$, i.e., $ \widetilde{\bm{H}}_{k,\text{DL}(\Delta)} = \widetilde{\bm{H}}_{k,\text{DL(p)}} - \widetilde{\bm{H}}_{k,\text{DL(TypeII)}}$, and a shortcut module is introduced to add the input $\widetilde{\bm{H}}_{k,\text{DL(TypeII)}}$ for acquiring the final output.
Besides, because the amplitude of $\widetilde{\bm{H}}_{k,\text{DL}(\Delta)}$ is generally so small that deep learning may be hard to accurately fit, a weighting coefficient $w$ is adopted to scale down the output value of deep learning, which can be empirically set around the average amplitude of $\widetilde{\bm{H}}_{k,\text{DL}(\Delta)}$.
In the corresponding output module, the FC layer transforms the feature vector from the pooling layer into the vectorized CSI difference, followed by the reshaping and weighting operations to obtain the ultimate reconstructed CSI difference $\widehat{\widetilde{\bm{H}}}_{k,\text{DL}(\Delta)}$.
%Concisely, the proposed output module can be summarized as
%\begin{equation}\label{eq17}
%\widehat{\widetilde{\bm{H}}}_{k,\text{DL(p)}} = ,
%\end{equation}

% Specifically, to adapt to the regression task, the FC layer of CSI reconstruction transforms tha feature vector into the vectorized CSI, followed by a reshaping operation to obtain the reconstructed CSI.

\subsection{Proposed Model Training with Two-Stage Loss}\label{S4.3}

For the CSI reconstruction task, the most widely adopted loss function is the mean squared error (MSE) \cite{ref16,ref17,ref18,ref19,ref20,ref21,ref22,ref23,ref24,ref25,ref26,ref27,ref28}, which can be written as
\begin{align}\label{eq17}
\text{loss}_\text{MSE} = \frac{1}{K N_\text{Tx} M} \sum_{k=1}^{K} \Big|\Big| \widetilde{\bm{H}}_{k,\text{DL(p)}} - \widehat{\widetilde{\bm{H}}}_{k,\text{DL(p)}} \Big|\Big|_\text{F}^2,
\end{align}
where $\widehat{\widetilde{\bm{H}}}_{k,\text{DL(p)}}$ denotes the reconstructed perfect angular-delay-domain CSI by deep learning.
Although the MSE loss can optimize the model parameters for approaching the perfect CSI label as close as possible, it may not obtain satisfactory performance in the MU-MIMO scenarios, which is exactly the key use case of the Type-II codebook \cite{ref36}.
Concretely, the reconstruction errors inevitably exist even for a well trained model.
Since the MSE loss does not take the interference coordination into account, the imperfect precoding based on the corresponding reconstructed CSI with errors is likely to impose considerable interference.
By contrast, the unsupervised learning method that uses the negative value of sum rate performance as the loss is usually adopted in the MU-MIMO scenarios \cite{ref32, ref33}, which could overcome the aforementioned drawback of MSE.
However, the mapping from the reconstructed CSI to the sum rate under the determined precoding method is a typical function from high dimensions to single dimension, which may possess massive local optimums that the negative sum rate loss would be stuck in.
This obviously restricts the performance of unsupervised learning methods.

Fortunately, it is noticed that the roles of the two loss functions above are complementary to some extent. Specifically, the gradients of MSE loss are devoted towards a clear global optimum, while the negative sum rate loss enjoys the capability of mitigating inter-UE interference.
Therefore, we propose a two-stage loss function that integrates the both advantages to enhance the sum rate performance in MU-MIMO scenarios.
In the first stage, the MSE loss in (\ref{eq17}) is adopted to implement a coarse reconstruction until its optimization has been almost sufficient (e.g., the decrease of MSE or the increase of sum rate is very slow).
Next, a weighted loss function including the MSE and negative sum rate is further applied during the second stage to alleviate the interference between different UEs under the guarantee of low reconstruction errors, which can be written as
\begin{align}\label{eq18}
\text{loss}_\text{stage2} = -R_\text{avg} + \mu \frac{1}{K N_\text{Tx} M} \sum_{k=1}^{K} \Big|\Big| \widetilde{\bm{H}}_{k,\text{DL(p)}} - \widehat{\widetilde{\bm{H}}}_{k,\text{DL(p)}} \Big|\Big|_\text{F}^2,
\end{align}
where $\mu$ is an adjustable weighting coefficient.

The practical deployment of the CSI reconstruction model is similar to the port selection model. It is noted that in order to support model training, the almost perfect CSI label can be acquired by channel estimation with long pilot sequences.

\section{Simulation Study}\label{S5}

\subsection{System Setup}\label{S5.1}

In the simulations, a single-BS single-sector FDD communication system with $K=5$ UEs is considered, where the 3GPP 38.901 channel model for urban macro-cell (UMa) scenarios is utilized to generate the CSI matrices \cite{ref42}.
Both LOS and NLOS scenarios are included in this channel model, where the LOS probability and corresponding pathloss are determined by the distance between BS and UE $d$ together with the UE height $h_\text{UE}$.
For simplicity, the BS is assumed to adopt zero-forcing precoding for simultaneously serving all UEs.
% Depending on the LOS/NLOS scenario, the channel specific model varies in the pathloss
For the R17 type-II codebook, the selection of $P = 32$ dominant angular-delay-domain ports from $N_\text{Tx} M = 256$ ports is considered for downlink channel measurement and feedback.
During the port coefficient compression at UE side, the wideband amplitude coefficient is quantized by $Q_\text{w} = 5$ bits, while $Q_\text{n,a}=3$-bit and $Q_\text{n,p} = 4$-bit quantizations are adopted for the narrowband amplitude and phase coefficients, respectively \cite{ref35}.
The detailed parameters of the simulated system and the Type-II codebook are listed in Table~\ref{tab1}.

% 38.901 Uma
% sector
% pathloss and los/nlos depends on h and d
% 5 users
% noise?(ignore now)
% polarization?(ignore now)

% In order to accurately simulate the inter-BS CSI dependence, we adopt the DeepMIMO dataset \cite{Ref:DATA} generated by the precise ray-tracing based on Wireless Insite \cite{Ref:WIRELESS_INSITE}. Specifically, the outdoor scenario `O1' shown in Fig.~\ref{fig4} is considered, where UE is randomly distributed in User Grid 1. The wireless HetNet consists of one low-frequency BS and $J=12$ mmWave BSs, where the low-frequency BS is located at BS $5$, and the mmWave BSs correspond to BS $1 \sim 12$. The default parameters of the simulated HetNet are listed in Table~\ref{tab1}. The power $\sigma^2$ of the AWGN for mmWave signals is calculated as $(-174 + 10\log_{10}W + N_{\text{F}})\,\text{dBm}$, while the AWGN power $\overline{\sigma}^2$ for low-frequency signals is calculated in the same way. The numbers of mmWave candidate narrow beams and wide beams are set to $N=64$ and $N/s = 16$, respectively.

\begin{table}[tp!] %table1
\vspace*{3mm}
\caption{System parameters.}
\label{tab1}
\vspace*{-5mm}
\begin{center}
\setlength{\tabcolsep}{2mm}
\begin{tabular}{ccccccc}
\toprule[0.8pt]
Parameters                                                                      & Values    \\ \toprule[0.8pt]
Uplink/downlink center frequency    $f_\text{c,UL}/f_\text{c,DL}$               & $3.4/3.5$ GHz \\
Number of UEs    $K$                                                             & $5$ \\
Sector range $\phi$                                                        & $120^\circ$ \\
% Downlink center frequency  $f_\text{c,DL}$                                    & $3.5$ GHz \\
Height of BS $h_\text{BS}$                                                      & $25$~m \\
Minimum/maximum heights of UE $h_\text{UE,min}/h_\text{UE,max}$                  & $1.5/22.5$~m \\
Minimum/maximum UE distances to BS $d_\text{min}/d_\text{max}$                   & $35/250$ m  \\
Maximum UE speed $v_\text{max}$                                                 & $3$ km/h \\
Downlink transmit power    $P_\text{Tx}$                                        & $35$ dBm  \\
% Noise power spectral density    $P_\text{Tx}$                                 & $-174$ dBm/Hz  \\
Noise factor      $N_{\text{F}}$                                                & $5$ dB \\
Number of BS antennas $N_\text{Tx}$                                             & $32$ \\
Numbers of BS horizonal/vertical antennas $N_\text{h}/N_\text{v}$               & $4/4$ \\
Subcarrier spacing         $f_\text{s}$                                         & $15$ kHz \\
Number of subbands         $M$                                                  & $8$ \\
Number of NLOS clusters  $N_\text{c,NLOS}$                                      & $12/20$ for LOS/NLOS scenarios \\
Number of paths within one NLOS cluster  $N_\text{p,NLOS}$                      & $20$  \\
Delay spread of clusters $\tau_\text{c,DS}$                                     & $98.3/406.5$~ns for LOS/NLOS scenarios \\
Angle spread of clusters $\theta_\text{c,AS}$                                   & $13.2^\circ/27.4^\circ$ for LOS/NLOS scenarios \\
Delay spread of paths within one cluster $\tau_\text{p,DS}$                     & $4.7$~ns\\
Angle spread of paths within one cluster $\theta_\text{p,AS}$                   & $5^\circ/2^\circ$ for LOS/NLOS scenarios \\
% Number of RBs in subband $N_\text{s}$                                         & 2 \\
% Number of total angular-delay-domain ports $N_\text{Tx} M$                    & $256$   \\
Number of selected angular-delay-domain ports $P$                               & $32$   \\
Quantization bit for wideband amplitude $Q_\text{w}$                            & $5$   \\
Quantization bits for narrowband amplitude/phase $Q_\text{n,a}/Q_\text{n,p}$    & $3/4$   \\
Oversampling factors of angular-domain horizonal/vertical bases $O_\text{h}/O_\text{v}$     & $1/1$ \\ \toprule[0.8pt]
\end{tabular}
\end{center}
\vspace{-5mm}
\end{table}

{
\begin{table}[tp!] %table2
\vspace{3mm}
\renewcommand\arraystretch{1.0}
\makeatletter\def\@captype{table}\makeatother\caption{Structures of proposed deep learning models.}
\label{tab2}
\vspace*{-5mm}
\begin{center}
\begin{tabular}{cccc}
\toprule[0.8pt]
\rule{0pt}{9pt}
\footnotesize{Modules}                                                     & \footnotesize{Layers} & \footnotesize{Parameters} \\
\toprule[0.8pt]
\multirow{6}{*}{\tabincell{c}{Convolutional \\ module}}
& Convolutional block 1                   & $f_\text{i}= 2, f_\text{o}= 256,\text{stride}=(1,1)$\\
& Convolutional block 2                   & $f_\text{i}= 256, f_\text{o}= 512, \text{stride}=(3,1)$\\
& Convolutional block 3                   & $f_\text{i}= 512, f_\text{o}= 512, \text{stride}=(3,3)$\\
& Convolutional block 4                   & $f_\text{i}= 512, f_\text{o}= 512, \text{stride}=(3,3)$\\
& Convolutional block 5                   & $f_\text{i}= 512, f_\text{o}= 512, \text{stride}=(3,3)$\\
% Pooling}}                                 & \multirow{2}{*}{\tabincell{c}{$f_{\text{i}}=512$, $f_\text{o}= 512$, max/average-pooling for \\ port selection/CSI reconstruction }} \\
& Pooling layer                                 & $f_{\text{i}}=512$, $f_\text{o}= 512$ \\
%                                                                          & FC                    & $f_{\text{i}}=1024,f_{\text{o}}=768,\text{dropout}=0.3,\text{softmax}$ \\
\toprule[0.8pt]
\multirow{2}{*}{\tabincell{c}{Output module for \\ port selection}}       & FC layer              & $f_\text{i}= 512, f_\text{o}= N_\text{Tx}M,\text{dropout}=0.3$  \\
                                                                          & Sigmoid function      & / \\
\toprule[0.8pt]
\multirow{2}{*}{\tabincell{c}{Output module for \\ CSI reconstruction}}    & FC layer              & $f_\text{i}= 512, f_\text{o}= N_\text{Tx}M,\text{dropout}=0.1$  \\
                                                                          & Shortcut              & / \\
\toprule[0.8pt]
\end{tabular}
\end{center}
\vspace{-5mm}
\end{table}
}

The specific structures and parameters of our deep learning based port selection and CSI reconstruction models are shown in Table~\ref{tab2}, where the both models adopt the same structure of the convolutional module for simplicity.
Concretely, $f_\text{i}$ and $f_\text{o}$ represent the numbers of input feature channels and output feature channels, respectively.
Besides, the two-dimensional convolutional kernel size $(3,3)$ and circular padding size $(1,1)$ as well as the LeakyReLU function with negative-axis slope $0.1$ are shared by all convolutional blocks, thus omitted in Table~\ref{tab2}.
Furthermore, the port selection model and CSI reconstruction model separately utilize the maximum pooling and average pooling in the pooling layer, and the dropout strategy \cite{Ref:DROPOUT} is applied in the FC layer for inhibiting overfitting.
For the port selection model, the focusing parameter in the focal loss is set to $\gamma = 2$ \cite{ref53}.

The training dataset containing $102,400$ samples and the validation dataset containing $2,560$ samples are constructed, respectively.
The deep learning models are trained for $200$ epochs with the learning rate $3 \times 10^{-6}$ during the training stage, where Adam optimizer based on the back propagation algorithm is used to optimize the model parameters \cite{Ref:ADAM}.
Since the initialization of deep learning models may influence the performance, the average results over the whole validation set with 3 training runs are ultimately used for evaluation.

\subsection{Port Selection}\label{S5.2}

In this subsection, the performance of port selection is evaluated by two metrics.
The first metric is the normalized CSI power of the selected angular-delay-domain ports $P_\text{N}$, which can be calculated as
%\begin{align}\label{eq19}
%P_\text{N} = \frac{ \sum_{p=1}^{P} \big|{\widetilde{\bm{H}}_\text{DL}}[\widehat{I}_{p}^{\text{(A)}},\widehat{I}_{p}^{\text{(D)}}]\big|^2 }{ \big|\big|\widetilde{\bm{H}}_\text{DL}\big|\big|_\text{F}^2 },
%\end{align}
\begin{align}\label{eq19}
P_\text{N} = \frac{ \sum_{p=1}^{P} \big|{\widetilde{\bm{H}}_\text{DL}}[{p}^{\text{(A)}},{p}^{\text{(D)}}]\big|^2 }{ \big|\big|\widetilde{\bm{H}}_\text{DL}\big|\big|_\text{F}^2 },
\end{align}
where ${p}^{\text{(A)}}$ and ${p}^{\text{(D)}}$ denote the angular-domain and delay-domain indices of the $p$-th selected port, respectively, and the UE index $k$ is omitted.
Clearly, higher $P_\text{N}$ indicates more accurate port selection with capturing larger CSI power.
The second metric is the average sum rate $R_\text{avg}$ defined in (\ref{eq4}).
For a fair comparison, we apply the standard procedure of R17 Type-II codebook except for port selection in this subsection.

\begin{figure}[tp!]
%\vspace{-6mm}
\begin{center}
\includegraphics[width=.8\textwidth]{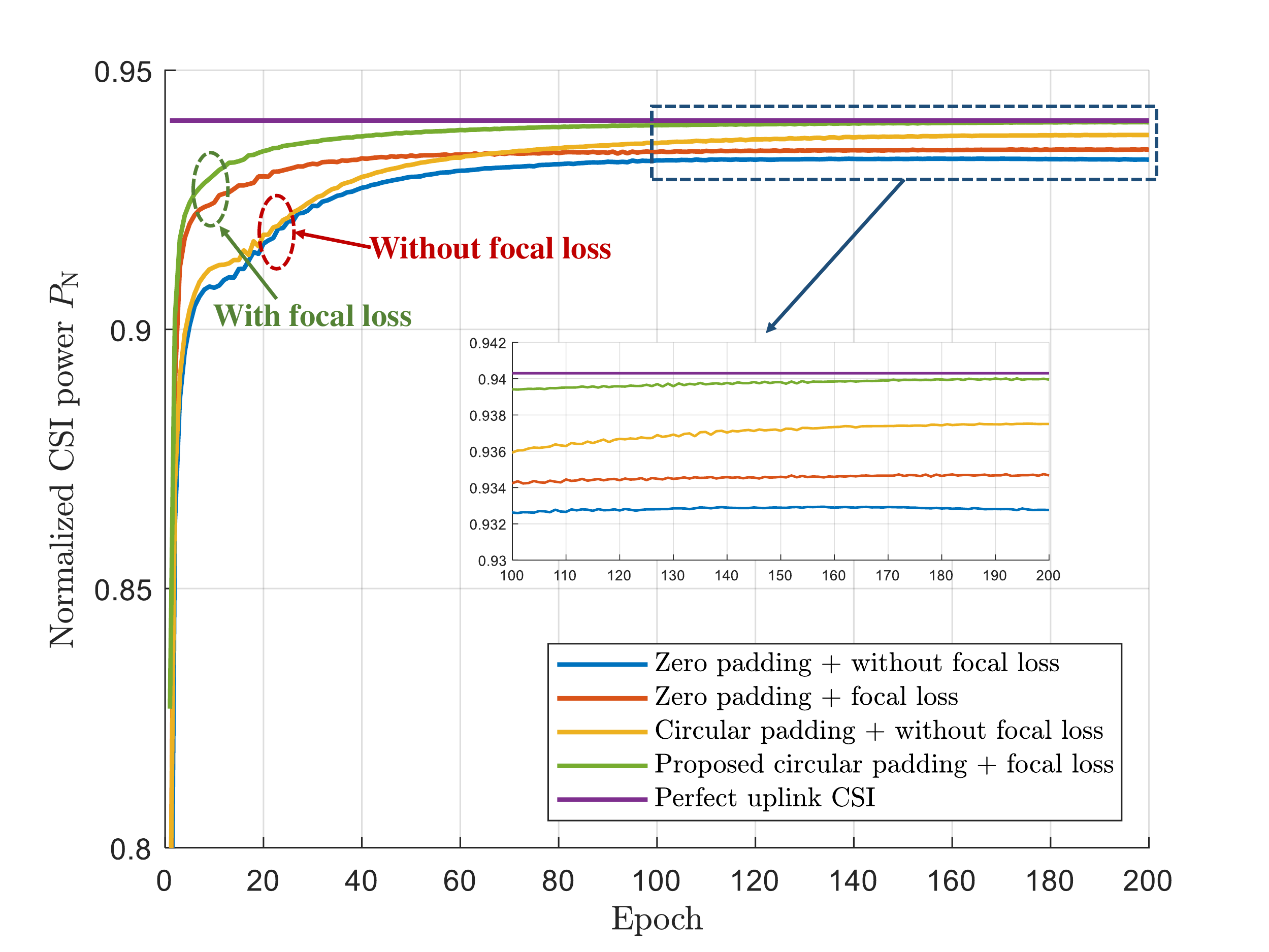}
\end{center}
\vspace{-8mm}
\caption{Convergence performance comparison for different deep learning designs in terms of normalized CSI power $P_\text{N}$, where uplink CSI SNR is set to $5$~dB and perfect uplink CSI without noise is illustrated as upper bound.}
\label{fig3}
\vspace{-6mm}
\end{figure}

Firstly, Fig.~\ref{fig3} depicts the convergence performance of our proposed deep learning model in terms of the normalized CSI power $P_\text{N}$ under the uplink CSI $\text{SNR}=5$~dB, where the impact of our adopted circular padding and focal loss is also investigated.
Specifically, two padding methods including conventional zero padding and our circular padding are compared, while whether or not to use the focal loss is considered in model training.
Besides, the $P_\text{N}$ performance of the port selection from perfect uplink CSI without noise is illustrated as the upper bound, which does not change with training epochs.
We can see that all the deep learning based methods have almost converged at around $100$-th epoch.
By comparing the converged $P_\text{N}$ of different deep learning designs, it can be concluded that both the circular padding and the focal loss are beneficial for achieving higher $P_\text{N}$, which validates the superiority of our proposed model design.
Besides, the models with the focal loss could obtain a significantly faster convergence speed before the $60$-th epoch, since the focal loss can concentrate on difficult samples for accelerating model optimization.
Overall, all the deep learning based methods can attain a very accurate port selection with $P_\text{N} > 93\%$, and our proposed model design with the circular padding and focal loss even achieves almost the same $P_\text{N}$ performance as the perfect uplink CSI.

\begin{figure}[tp!]
%\vspace{-6mm}
\begin{center}
\includegraphics[width=.8\textwidth]{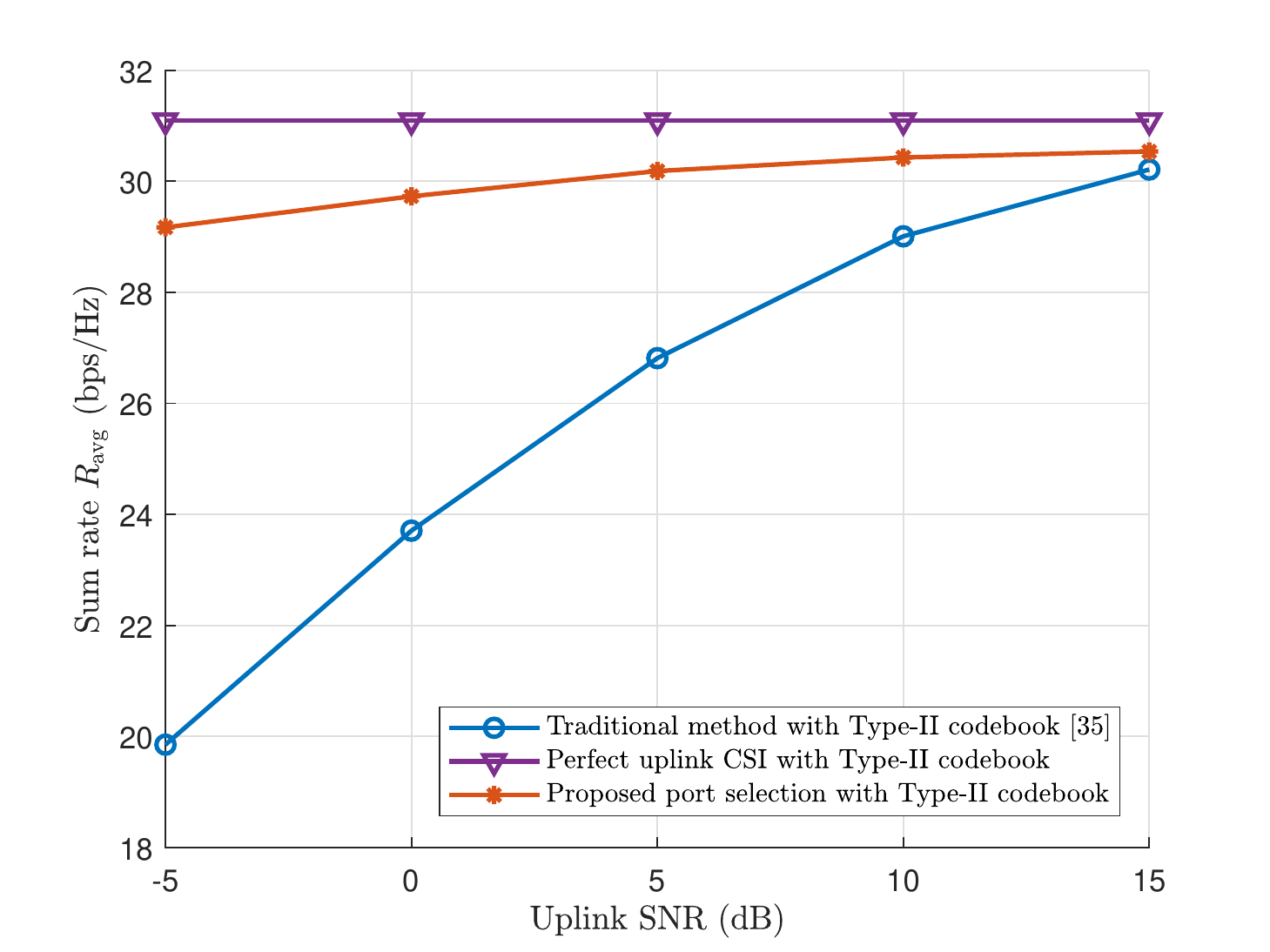}
\end{center}
\vspace{-8mm}
\caption{Sum rate performance comparison for different port selection methods as function of uplink SNR, where perfect uplink CSI without noise is illustrated as upper bound.}
\label{fig4}
\vspace{-6mm}
\end{figure}

Next, the sum rate performance $R_\text{avg}$ of our proposed port selection method and the Type-II codebook \cite{ref35} under different SNRs of uplink CSI are compared in Fig.~\ref{fig4}, where the achievable $R_\text{avg}$ of the perfect port selection based on the uplink CSI without noise is also depicted.
It can be seen that the sum rate performance $R_\text{avg}$ of our proposed method surpasses the Type-II codebook in all evaluated SNRs from $-5~\text{dB}$ to $15~\text{dB}$.
Especially for the severely low-SNR scenarios, the proposed method can improve the achievable $R_\text{avg}$ of the Type-II codebook by more than $40\%$.
Besides, our proposed method enjoys the more robust $R_\text{avg}$ performance to the uplink SNR than the Type-II codebook, and only suffers from $6.2\%$ rate loss compared to the perfect uplink CSI in the scenarios with $-5~\text{dB}$ uplink SNR.
This demonstrates that our proposed deep learning based port selection method is capable of well adapting to different channel environments with various SNRs.

\begin{figure}[tp!]
%\vspace{-6mm}
\begin{center}
\includegraphics[width=.8\textwidth]{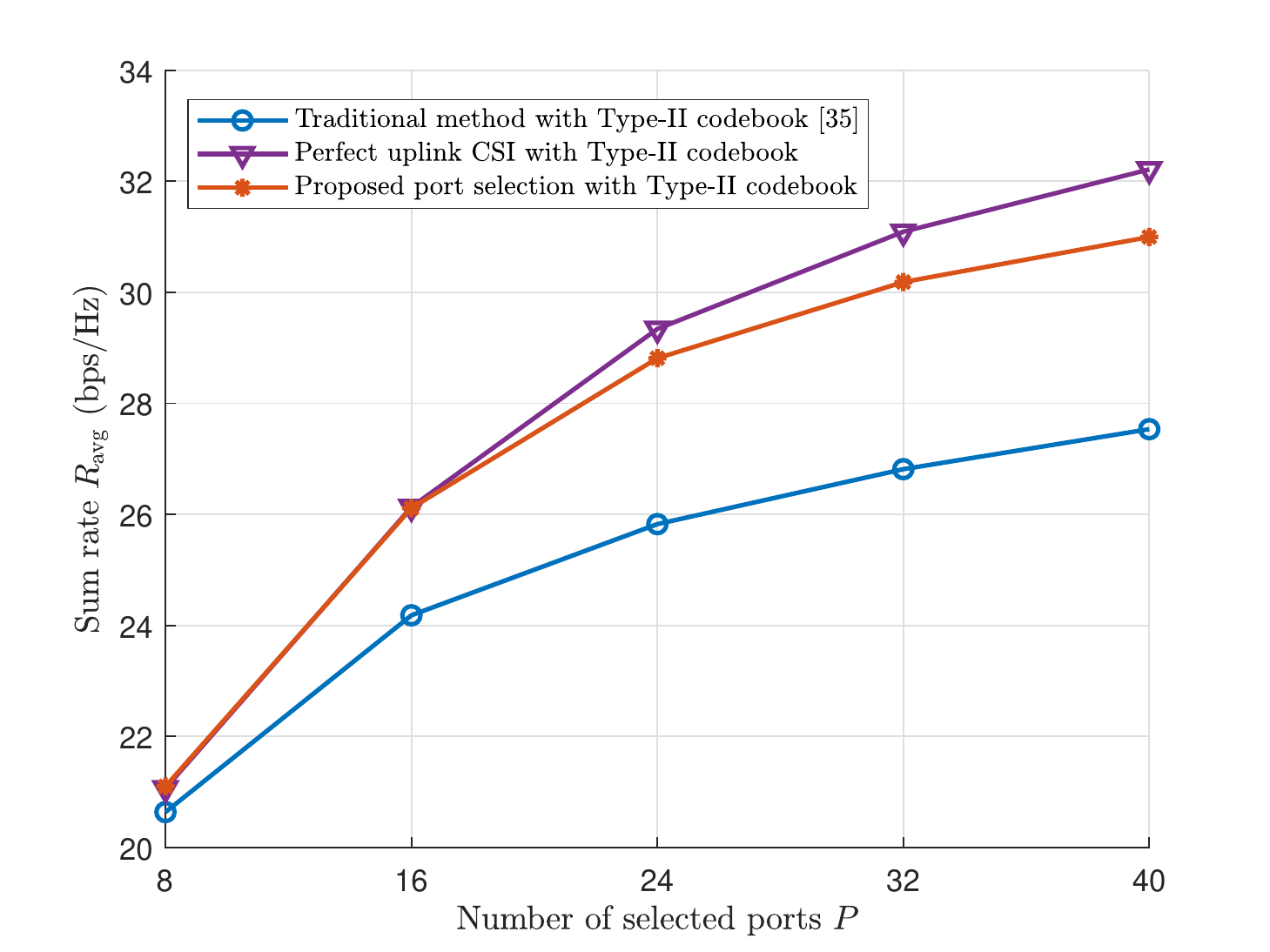}
\end{center}
\vspace{-8mm}
\caption{Sum rate performance comparison for different port selection methods as function of selected port number $P$, where uplink CSI SNR is set to $5$~dB and perfect uplink CSI without noise is illustrated as upper bound.}
\label{fig4.5}
\vspace{-6mm}
\end{figure}

Furthermore, Fig.~\ref{fig4.5} illustrates the sum rate $R_\text{avg}$ of our port selection method and the Type-II codebook as the function of selected port number $P$, where the SNR of uplink CSI is set to $5$~dB and the perfect uplink CSI without noise is depicted as the upper bound.
It is evident that the achievable $R_\text{avg}$ of all port selection methods increases with $P$ owing to capturing more CSI power.
Then, we can see that our deep learning based method could attain more remarkable $R_\text{avg}$ enhancement over the Type-II codebook as $P$ increases, since larger $P$ indicates the inclusion of more relatively weak ports that the traditional Type-II codebook are hard to accurately choose.
Besides, the $R_\text{avg}$ performance of our proposed method approaches the perfect uplink CSI without noise.
In particular, our proposed method enjoys almost the same $R_\text{avg}$ to the perfect uplink CSI at $P = 8$ and $16$.

\subsection{CSI Reconstruction}\label{S5.3}

Based on the proposed port selection method, we further investigate the CSI reconstruction performance in terms of the average sum rate $R_\text{avg}$.
For our proposed two-stage loss function, in order to ensure sufficient model optimization by the MSE loss in the first stage, the criterion of switching to the second stage is set as the increase of $R_\text{avg}$ smaller than $0.02$ for $5$ consecutive epochs.

\begin{figure}[tp!]
%\vspace{-6mm}
\begin{center}
\includegraphics[width=.8\textwidth]{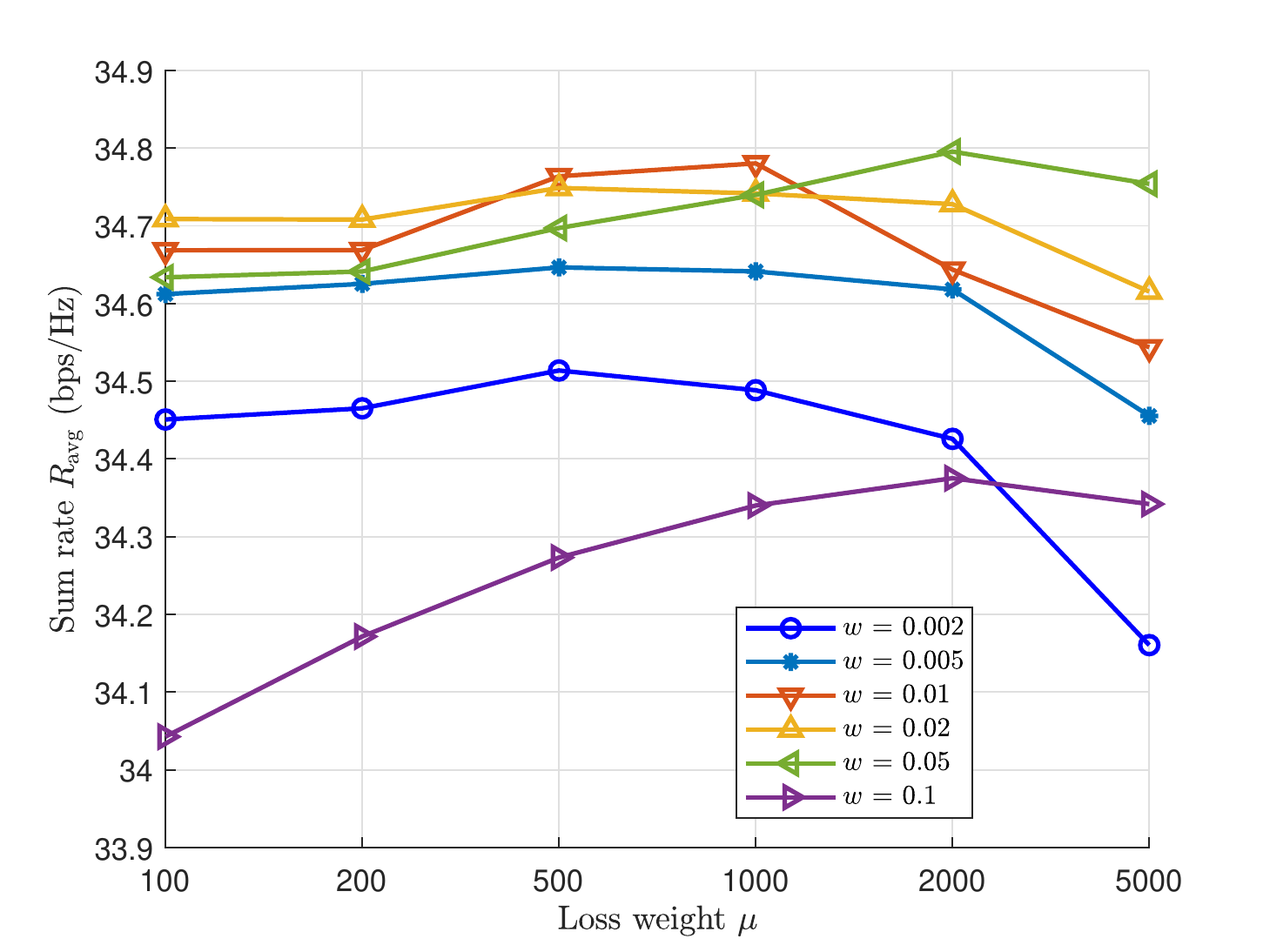}
\end{center}
\vspace{-8mm}
\caption{Sum rate performance comparison under different weighting coefficients in shortcut module $w$ and loss function $\mu$, where uplink SNR is set to $5~\text{dB}$.}
\label{fig5}
\vspace{-6mm}
\end{figure}

Firstly, we investigate the impact of the weighting coefficients in our proposed shortcut module $w$ and two-stage loss function $\mu$ on the sum rate performance $R_\text{avg}$ in Fig.~\ref{fig5}, where the SNR of uplink CSI is set to $5$~dB.
Since the average amplitude of the model fitting target, i.e., the angular-delay-domain CSI difference $\widetilde{\bm{H}}_{\text{DL}(\Delta)}$, is $0.017$ in the training dataset, the empirical value of $w$ is chosen from a nearby range $\mathcal{W} = \{0.002,0.005,0.01,0.02,0.05,0.1\}$.
% It can be seen that both overwhelmingly small or large $w$ could degrade the achievable $R_\text{avg}$, while the moderate $w \in \{ 0.01,0.02,0.05 \}$ around $0.017$ enjoys the robust high sum rate performance with $R_\text{avg} = 34.7~\text{bps/Hz}$.
It can be seen that both overwhelmingly small and large $w$ could degrade the achievable $R_\text{avg}$, while the moderate $w \in \{ 0.01,0.02,0.05 \}$ around $0.017$ enjoys the robust high sum rate performance.
For these proper $w$, the loss weight $\mu$ has slight influence on the sum rate, where the stable performance around $R_\text{avg} = 34.7~\text{bps/Hz}$ could be guaranteed from $\mu=200$ to $2000$.
This is because the MSE loss in the second stage is mainly utilized for constraining the optimization direction towards the perfect CSI label, which can be well satisfied within a relatively wide range of $\mu$.
% These results demonstrate that the two-stage loss function is a relatively robust design to the hyper-parameters.
% This result reflects the intrinsic consistency of the basic optimization directions between the MSE loss and the negative sum rate loss, so that the MSE loss is mainly to constraint the optimization  and .
Based on Fig.~\ref{fig5}, we adopt the optimal parameter configuration $w=0.05,\mu=2000$ in the following simulations.

\begin{figure}[tp!]
%\vspace{-6mm}
\begin{center}
\includegraphics[width=.8\textwidth]{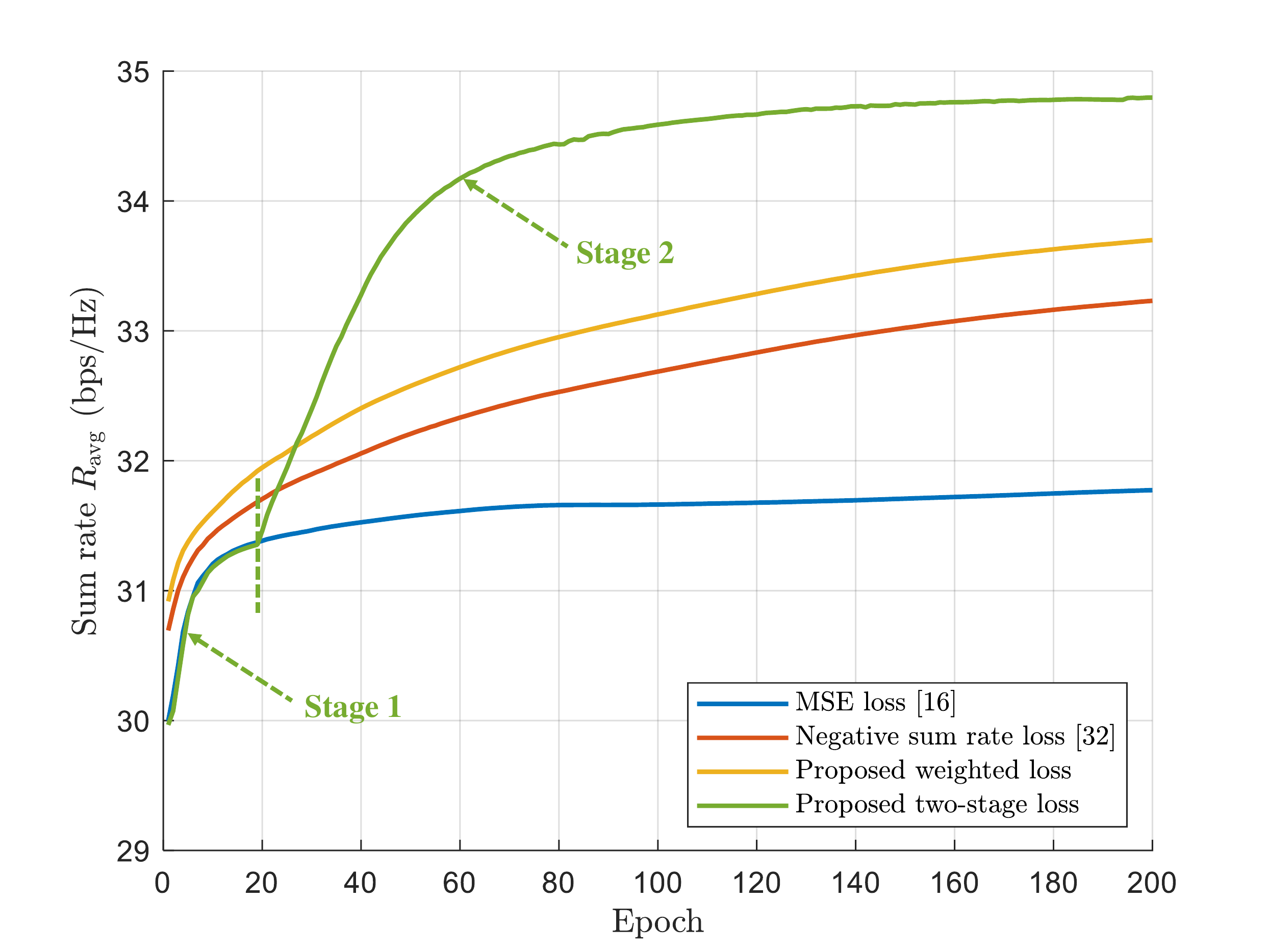}
\end{center}
\vspace{-8mm}
\caption{Sum rate performance comparison for different loss functions, where uplink SNR is set to $5~\text{dB}$.}
\label{fig6}
\vspace{-6mm}
\end{figure}

Next, Fig.~\ref{fig6} compares the convergence performance of different loss functions in terms of the sum rate performance $R_\text{avg}$ under the uplink $\text{SNR}=5~\text{dB}$, where the conventional MSE loss \cite{ref16} and negative sum rate loss \cite{ref32} are adopted as our baselines.
Apart from the proposed two-stage loss, we further consider its simplified counterpart that applies the weighted loss in (\ref{eq18}) throughout model training.
It is obvious that both of our proposed weighted loss and two-stage loss outperform the conventional MSE and negative sum rate losses.
Besides, our proposed two-stage loss attains about $34.6~\text{bps/Hz}$ at the $100$-th epoch, which enjoys more prompt convergence and higher $R_\text{avg}$ performance than the weighted loss.
This is because the coarse alignment to the perfect CSI based on the MSE loss in the first stage could effectively guide the negative sum rate loss to find a better optimum and avoid being trapped in the local optimums at the beginning.

\begin{figure}[tp!]
%\vspace{-6mm}
\begin{center}
\includegraphics[width=.8\textwidth]{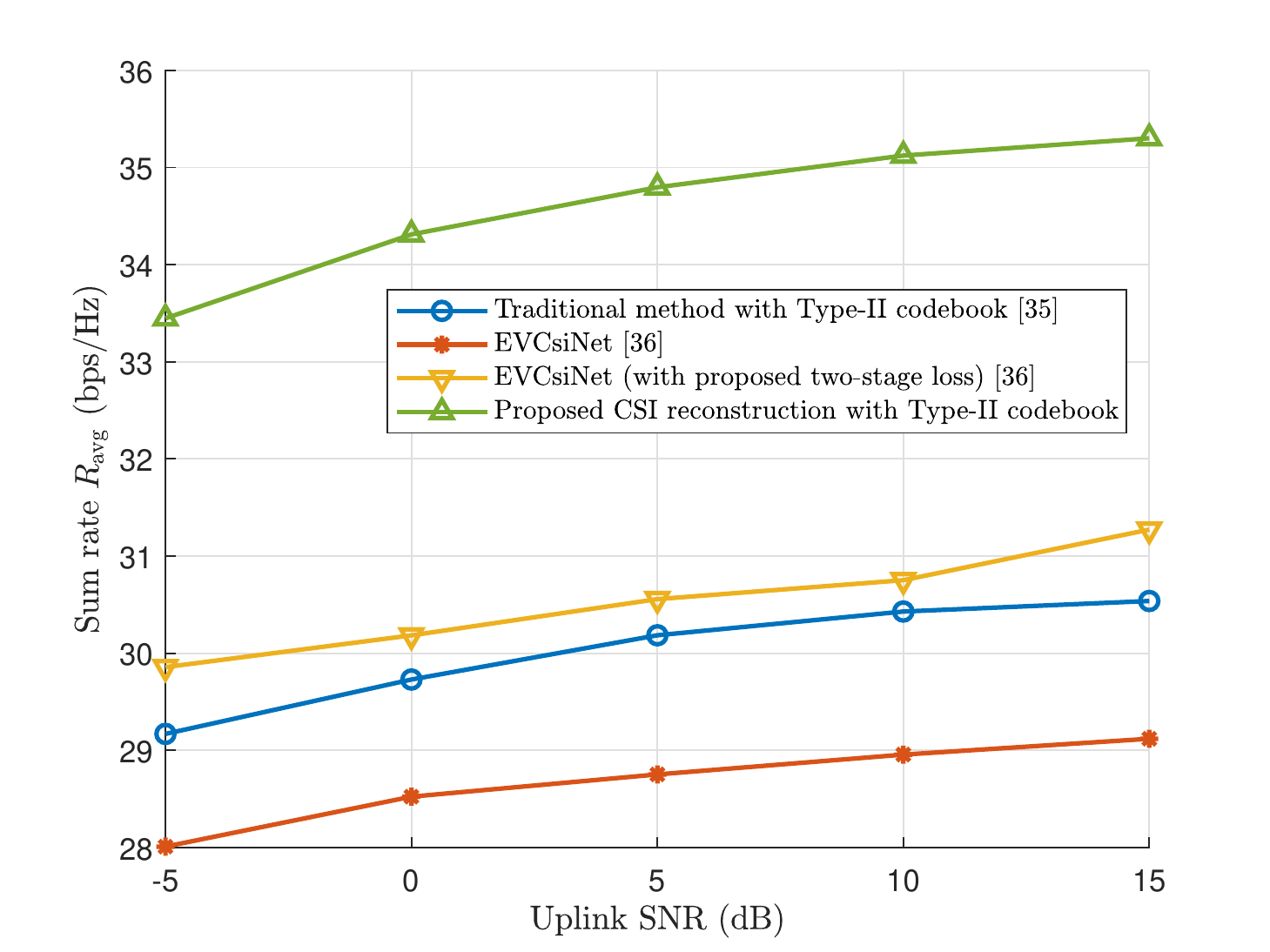}
\end{center}
\vspace{-8mm}
\caption{Sum rate performance comparison for different CSI feedback methods as function of uplink SNR.}
\label{fig7}
\vspace{-6mm}
\end{figure}

Figure \ref{fig7} further compares the sum rate performance $R_\text{avg}$ of different CSI feedback methods as the function of uplink CSI SNR, where the traditional feedback procedure of R17 Type-II codebook \cite{ref35} and the AE based EVCsiNet for the R16 Type-II codebook \cite{ref36} are adopted as our baselines.
For a fair comparison, the number of feedback bits in EVCsiNet is equal to the R17 Type-II codebook, and the EVCsiNet trained by our proposed two-stage loss is also simulated.
To effectively enhance the performance of EVCsiNet, the adaptive stacked sigmoid approximation is adopted during the gradient back propagation to facilitate fitting the quantization process \cite{ref23}, and \emph{we further rearrange the channel coefficients by their corresponding delay-domain and angular-domain indices to reserve the local sparse structures among selected neighboring ports}.
It can be seen that the $R_\text{avg}$ performance of the conventional EVCsiNet is poorer than the standard Type-II codebook, which validates that deep learning is not expert in compressing the non-sparse vector of port coefficients.
By contrast, our proposed CSI reconstruction method could achieve significantly higher $R_\text{avg}$ than all the baselines, which demonstrates that our proposed method could effectively leverage the sparse structure information at BS side for improving the accuracy of CSI reconstruction.

\begin{figure}[tp!]
\vspace{-3mm}
\begin{center}
\includegraphics[width=.8\textwidth]{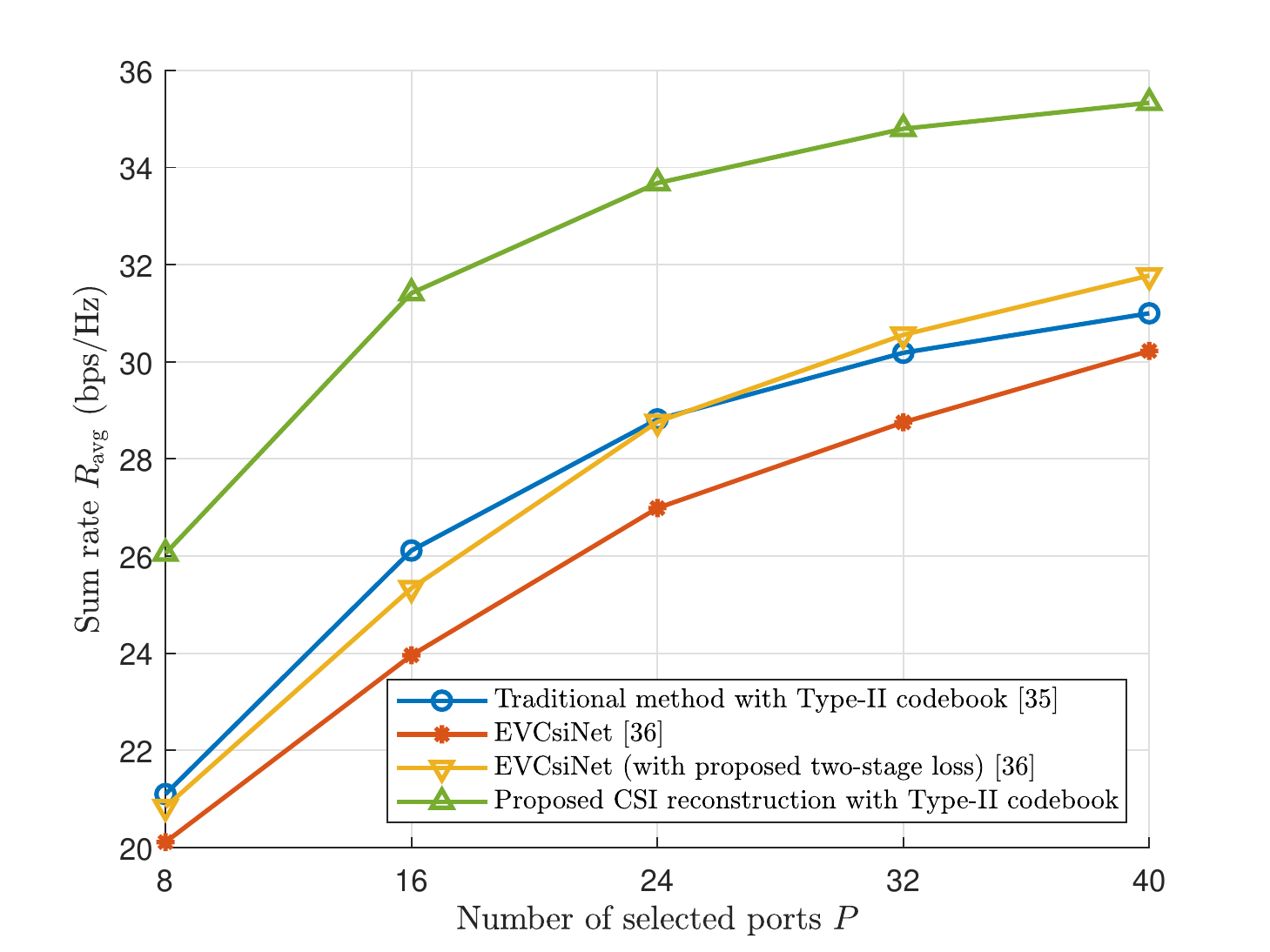}
\end{center}
\vspace{-8mm}
\caption{Sum rate performance comparison for different CSI feedback methods as function of selected port number $P$, where uplink SNR is set to $5~\text{dB}$.}
\label{fig7.5}
\vspace{-6mm}
\end{figure}

Finally, the sum rate performance $R_\text{avg}$ of our CSI reconstruction method and the baselines under different numbers of selected ports $P$ is depicted in Fig.~\ref{fig7.5}, given the uplink CSI $\text{SNR}=5$~dB.
We can see that our proposed method can improve the achievable $R_\text{avg}$ by more than $10\%$ over the Type-II codebook and the enhanced EVCsiNet with our two-stage loss at all evaluated $P$, which again verifies the superiority of our CSI reconstruction method.

\section{Conclusions}\label{S6}

In this paper, a new paradigm is proposed to improve the CSI feedback performance of R17 Type-II codebook.
Firstly, deep learning is leveraged to accurately select the dominant angular-delay-domain ports from the uplink CSI with relatively low SNRs.
Specifically, the port selection problem is formulated as a multi-label classification task, where the circular-padding based CNN architecture is designed to facilitate the extraction of the sparse structure, while the focal loss is harnessed to solve the class imbalance issue.
Secondly, the standard feedback in the R17 Type-II codebook is used to reconstruct the CSI eigenvectors by deep learning at BS side to exploit the sparse features, where a weighted shortcut module is utilized to enhance the reconstruction accuracy.
Finally, a two-stage loss function integrating the MSE loss and the negative sum rate loss is designed to effectively reconstruct the downlink CSI and mitigate the inter-UE interference simultaneously.
The simulation study demonstrates that our proposed deep learning based port selection and CSI reconstruction methodology could enhance the sum rate performance by more than $10\%$ when compared to its conventional R17 Type-II codebook and deep learning benchmarks.

{
}

\end{document}